\documentclass[sigconf]{aamas}

\usepackage{balance} 
\usepackage{blkarray}
\usepackage{algorithm}
\usepackage{algorithmic}
\usepackage{float}
\usepackage{multirow} 
\usepackage{csquotes}

\usepackage{xr}
\makeatletter

\newcommand*{\addFileDependency}[1]{
\typeout{(#1)}
\@addtofilelist{#1}
\IfFileExists{#1}{}{\typeout{No file #1.}}
}\makeatother

\newcommand*{\myexternaldocument}[1]{%
\externaldocument{#1}%
\addFileDependency{#1.tex}%
\addFileDependency{#1.aux}%
}

\myexternaldocument{appendix}

\setcopyright{ifaamas}
\acmConference[AAMAS '25]{This paper is accepted by the 24th International Conference
on Autonomous Agents and Multiagent Systems (AAMAS 2025) as a Full Paper}
\copyrightyear{}
\acmYear{}
\acmDOI{}
\acmPrice{}
\acmISBN{}

\acmSubmissionID{<<OpenReview submission id>>}

\title[AAMAS-2025 Full]{Bottom-Up Reputation Promotes Cooperation with \\ Multi-Agent Reinforcement Learning}

\author{Tianyu Ren } 
\orcid{0000-0002-7053-2796}
\affiliation{
  \institution{University of Manchester}
  \city{Manchester}
  \country{United Kingdom}}
\email{tianyu.ren@manchester.ac.uk}

\author{Xuan Yao}
\orcid{0000-0002-8549-2096}
\affiliation{
  \institution{Southeast University}
  \city{Nanjing}
  \country{China}}
\email{shiny\_yao@yeah.net}

\author{Yang Li}
\orcid{0009-0002-7073-9698}
\affiliation{
  \institution{University of Manchester}
  \city{Manchester}
  \country{United Kingdom}}
\email{yang.li-4@manchester.ac.uk}

\author{Xiao-Jun Zeng}
\orcid{0000-0002-2320-2495}
\affiliation{
  \institution{University of Manchester}
  \city{Manchester}
  \country{United Kingdom}}
 \email{x.zeng@manchester.ac.uk}

\begin{abstract}
Reputation serves as a powerful mechanism for promoting cooperation in multi-agent systems, as agents are more inclined to cooperate with those of good social standing. While existing multi-agent reinforcement learning methods typically rely on predefined social norms to assign reputations, the question of how a population reaches a consensus on judgement when agents hold private, independent views remains unresolved. In this paper, we propose a novel bottom-up reputation learning method, \textbf{\textit{L}}earning with \textbf{\textit{R}}eputation \textbf{\textit{R}}eward (\textbf{\textit{LR2}}), designed to promote cooperative behaviour through rewards shaping based on assigned reputation. Our agent architecture includes a dilemma policy that determines cooperation by considering the impact on neighbours, and an evaluation policy that assigns reputations to affect the actions of neighbours while optimizing self-objectives. It operates using local observations and interact- ion-based rewards, without relying on centralized modules or predefined norms. Our findings demonstrate the effectiveness and adaptability of LR2 across various spatial social dilemma scenarios. Interestingly, we find that LR2 stabilizes and enhances cooperation not only with reward reshaping from bottom-up reputation but also by fostering strategy clustering in structured populations, thereby creating environments conducive to sustained cooperation.
\end{abstract}


\newcommand{\BibTeX}{\rm B\kern-.05em{\sc i\kern-.025em b}\kern-.08em\TeX}

\begin{document}


\pagestyle{fancy}
\fancyhead{}


\maketitle 


\section{Introduction}
Cooperation is ubiquitous in almost every form of social interaction and is arguably a key factor in the success of complex social systems, such as economics~\cite{zheng2022ai}, evolutionary biology~\cite{sigmund2010social}, and multi-agent systems (MAS)~\cite{conitzer2023foundations}.  With the rise of artificial intelligence (AI), autonomously operating learning agents within MAS are becoming increasingly common. A typical example is autonomous vehicles, which must share the road with both other vehicles and human drivers; without proper coordination, issues such as road congestion may arise~\cite{liang2022federated}. However, achieving effective collective actions among self-interested agents remains a significant challenge. 

To address this challenge, mechanisms that facilitate correlated interactions among autonomous, decentralized learning agents are required~\cite{fatima2024learning}. These mechanisms shape the interaction structures or rewards within a population to promote cooperation over defection. In evolutionary biology and economics, indirect reciprocity (IR) and direct reciprocity (DR) have been identified as key drivers of cooperation~\cite{nowak2006five}. Notably, IR—offering a compelling explanation for the evolution of cooperative behaviour among MAS agents~\cite{nowak2005evolution}—differs from DR, where benefits are received directly from those assisted~\cite{van2012direct}; instead, IR allows individuals to benefit from the broader community. Consequently, sustaining cooperation via IR necessitates social information reflecting the historical behaviours of agents, which is often governed by social norms. These norms establish expected patterns and may impose penalties for violations. Enforcement typically involves rumour~\cite{kawakatsu2024mechanistic} or reputation~\cite{ohtsuki2004should}, which assess the \enquote{goodness} of agents and enable selective altruism, where individuals help those with good reputations.

Studies in MAS have demonstrated the importance of IR by incorporating it into traditional multi-agent reinforcement learning (MARL) tasks~\cite{vinitsky2023learning,xu2019cooperation}. Reputation allows agents to reshape rewards by assigning bonuses or penalties based on observed social information. For instance, agents can integrate moral values~\cite{tennant2023modeling} and normative punishment~\cite{vinitsky2023learning} to enforce compliance and learn behavioural rules. Reputation also serves as an informative signal for optimizing behaviour~\cite{anastassacos2021cooperation} or forming selective interaction relationships~\cite{ren2024enhancing,mckee2023scaffolding} across different scenarios. This body of research not only promotes cooperation and explains the formalization of artificial ethics and morality among MAS but also provides a feasible pathway for constructing a normative human-AI society.

Despite promising developments, reputation-based strategies and norms in MARL face several challenges. Standard IR theory assumes that reputations are common knowledge, meaning the entire population agrees on the evaluation process~\cite{santos2021complexity}. This consensus stabilizes reputation dynamics and promotes cooperation by setting clear expectations. However, when agents form private assessments, disagreements can arise, leading to perceptions of unfairness that may undermine cooperation~\cite{kawakatsu2024mechanistic}. Constraining the norm space through structured interactions appears essential to prevent cooperative collapse~\cite{murase2024computational}. Moreover, while social norms are often enforced exogenously by central institutions, they vary widely across different environments and populations~\cite{kessinger2023evolution}. Establishing appropriate norms thus demands extensive prior knowledge of the environment—a formidable challenge in complex MARL settings. These issues reduce the adaptability of reputation mechanisms and raise two fundamental questions: To what extent do individuals share a consensus on reputations, and how are social norms used to evaluate and assign reputations?

In this paper, we address these problems by developing Learning with Reputation Reward (LR2), a training method that promotes cooperation through reshaped rewards based on bottom-up reputation. Each agent has two policies: a dilemma policy for determining action and an evaluation policy for assigning reputations. LR2 agents aim both to maximise personal benefits without compromising their reputation and to assign reputations that incentivise prosocial behaviour within their local group while furthering their own interests. It reshapes the static reward function in a distributed manner by integrating assigned reputations from neighbours. Unlike similar reputation-based MARL methods, our approach emphasises the heterogeneity and endogeneity of norm formation, framing reputation assignment as a learning process rather than a reactive adaptation to predefined social norms~\cite{smit2024learning,oldenburg2024learning,xu2019cooperation}. This approach captures the coevolution of reputation and cooperation, offering a pathway to predict whether, and by what dynamics, a population will achieve coordination under IR. We validate LR2 in various social dilemmas, showing that it not only influences opponents' behavioural strategies through reputation assignment but also promotes strategy clustering in structured populations. Furthermore, we demonstrate that biased social information—where peers assign differing reputations to the same individual—can either facilitate or hinder cooperation, depending on the dilemma's strength. In summary, our work makes three contributions:
\begin{itemize}
    \item We propose a novel training method LR2, which promotes cooperation through bottom-up reputation. It enables agents to consider their impact on others while optimizing their objectives with learned reputation assignments.
    \item We reveal how LR2 promotes cooperation by examining the coevolution of cooperation and reputation. LR2 stabilizes prosocial behaviour through emerging normative judgments and further enhances cooperation via strategy clustering.
    \item We demonstrate that LR2 overcomes the scalability limitations of predefined norms in various social dilemma scenarios. It enables agents to learn and converge on shared norms in a decentralized and sample-efficient manner.
\end{itemize}

\section{Related Works}

\textit{Cooperation in MARL.} MARL in collaborative settings can be dichotomized into two branches: team-based and mixed-motive environments~\cite{du2023review}. In the former, agents coordinate actions and share a single scalar reward. Several recent advanced methods have been proposed to solve this problem~\cite{yu2022surprising,sunehag2018value,son2019qtran}. By contrast, in mixed-motive environments, agents receive individual rewards, necessitating a balance between personal maximisation and social welfare. Such systems often encounter social dilemmas~\cite{van2013psychology}, where individually rational decisions lead to collectively suboptimal outcomes. In MAS, these dilemmas extend spatially and temporally, forming sequential social dilemmas~\cite{leibo2017multi}. A core challenge is understanding how cooperation among self-interested agents can emerge and remain stable, despite threats like conflict, overconsumption, free-riding, and defection~\cite{du2023review}. Previous research has explored solutions through other-regarding preferences~\cite{mckee2021multi,hughes2018inequity}, reputation mechanisms~\cite{anastassacos2021cooperation,smit2024learning}, and anticipation of future behaviours~\cite{foerster2018learning,yang2020learning}. Notably, reputation serves as an adaptable measure of social standing that can guide subsequent agent interactions~\cite{anastassacos2020partner}. Building on this, our paper focuses on encouraging cooperation in MARL through the auxiliary learning dynamics inherent in reputation assignments.

\textit{Reputation and Social Norms.} In the IR mechanism, actions are evaluated against social norms that integrate behavioural and reputational information to update an agent's standing~\cite{nowak2005evolution}. These norms can be imposed top-down by a central authority or emerge bottom-up through interactions~\cite{savarimuthu2011norm}. In the top-down approach, norms are designed offline and uniformly applied~\cite{santos2016social}. For instance, using Boo- lean inputs for reputation and strategy, norms can be encoded as a 4-bit string, $d=(d_{G, C},d_{G, D},d_{B, C},d_{B, D})_2$. Then, agents update reputations based on second-party reports with imperfect observations~\cite{haynes2017engineering}. Since reputations are repeatedly evaluated with small errors, this process can be modelled as an ergodic Markov chain with a characterised stationary distribution~\cite{ohtsuki2004should}, determining the fitness of an agent’s rule $\pi_i$. Previous studies indicate that the well-known norm Stern Judging (SJ) effectively drives behavioural dynamics through both imitation processes~\cite{santos2016social} and Reinforcement Learning (RL)~\cite{smit2024learning}. SJ assigns good reputations to agents who cooperate with good partners and defect against bad ones, while assigning bad reputations to those who act the opposite.

\textit{MARL and Bottom-Up Norm.} In MARL, the centralized training with decentralized execution (CTDE) framework~\cite{lowe2017multi} lacks centralized information for top-down judgment during execution. A bottom-up approach, however, enables agents to adopt decentralized methods for assigning reputations, allowing norms to emerge organical- ly~\cite{xu2019cooperation}. Recent studies explore how agents learn to assign and respond to reputations. These norms can be learned either through explicit payoff-based methods~\cite{xu2019cooperation} or by inferring the existence of shared normativity via approximate Bayesian rule induction~\cite{oldenburg2024learning}. Beyond reputation, the bottom-up creation of norms can also manifest as intrinsic rewards~\cite{mckee2021multi} or public sanctions~\cite{vinitsky2023learning}. However, maintaining cooperation under learned norm-guided behaviour is constrained by the presence of a predefined set of norms. In this paper, we extend this approach by allowing agents to assign reputations to their interacting neighbours based solely on their own rewards, which does not require knowledge of a predefined set of norms beyond the agents' observations. 

\section{Preliminaries}
\subsection{Social Dilemmas}
We examine social dilemmas using a spatial model of repeated two-player symmetric games, where agents interact with neighbours by uniformly choosing to cooperate ($C$) or defect ($D$). The game is defined by the payoff matrix $M$:
\begin{equation} \label{eq:payoff_matrix}
    M= \begin{blockarray}{cccc}
        & C & D\\
      \begin{block}{c(ccc)}
        C & R,R & S,T \\
        D & T,S & P,P  \\ 
      \end{block}
    \end{blockarray}.
\end{equation}

Here, mutual cooperation yields reward $R$ and mutual defection yields punishment $P$. In asymmetric cases, where one agent cooperates and the other defects, the cooperator receives the sucker's payoff $S$ and the defector obtains the temptation payoff $T$. These payoffs define the social dilemma~\cite{macy2002learning}:

\begin{enumerate}
    \item Snowdrift Game (SG): $T > R > S > P$. The optimal strategy depends on the opponent's action, making unilateral cooperation advantageous.
    \item Stag-Hunt Game (SH): $R > T > P > S$. Unilateral cooperation leads to a loss if the other player defects, emphasizing the importance of mutual cooperation.
    \item Prisoner's Game (PD): $T > R > P > S$. Mutual defection is the equilibrium outcome, even though mutual cooperation offers a better payoff for both players.     
\end{enumerate}

The PD, which blends aspects of SG and SH, poses the greatest challenge for fostering cooperation. Following the convention~\cite{santos2006evolutionary} of normalising the difference between $R$ and $P$ to $1$, we set $R = 1$ and $P = 0$, with $T$ and $S$ constrained to $0\leq T\leq 2$ and $-1\leq S \leq 1$, respectively.

\subsection{Reputation and Social Norms}
Reputation is public information shared among neighbouring agents, derived from their strategies to measure cooperativeness. Unlike previous studies using binary reputation values~\cite{anastassacos2021cooperation,podder2021local}, we model reputation as a continuous attribute ranging from $0$ to $1$, indicating an agent's degree of \enquote{Good} (G) or \enquote{Bad} (B). This extension enhances the robustness of agent learning by capturing finer variations in social standing regarding agent behaviour. Initially, the G and B labels are nominal with no inherent meaning; their significance emerges through agents' actions during the evolution of the game. An agent's decision to cooperate depends on its current reputation and the observed reputations of its neighbours. In pairwise donor games with binary reputations, common strategies include always cooperate (ALLC), always defect (ALLD), and discriminate (DISC), where DISC agents cooperate with good-reputation recipients and defect against bad ones~\cite{santos2016social}. However, since reputation in our model is continuous and agents interact with groups of neighbours rather than single recipients, the strategy representations used in previous studies~\cite{santos2018social} must be expanded. Specifically, the strategy tuple $\Pi$ becomes an infinitely large bit string $\Pi = (\pi_1,\pi_2,\dotsc)$, where each $\pi \in \{0,1\}$ represents whether an agent cooperates for a given combination of self and neighbour reputations. We employ a policy network to specify the behaviour rules for each agent, as detailed in Section~\ref{Sec: Method_Repu}. 

\subsection{Markov Games}
We formalize the evaluation process in social dilemmas as an $N$-agent, partially observed, general-sum Markov games (POMGs)~\cite{littman1994markov,liu2022sample,perolat2017multi}, augmented with the concept of bottom-up reputation and an evaluation function that assesses the behaviours of neighbourhoods. A POMG is given by the tuple $M=\langle \mathcal{S,A,O,T,R,\gamma} \rangle$, where $\mathcal{A}\coloneq\mathcal{A}^1\times\cdots\times\mathcal{A}^N$ and $\mathcal{O}\coloneq\mathcal{O}^1\times\cdots\times\mathcal{O}^N$ denotes the joint action observations and the joint action space of all $N$ agents, indexed by $i\in[N]\coloneq\{1,\dotsc,N\}$. The game operates on a finite set of states $\mathcal{S}$, with each agent's $d$-dimensional observation of the state mapped by the function $\mathbb{O}: S\times\{1,\dotsc,N\}\rightarrow\mathbb{R}^d$. In each state, the agent $i$ selects an action from its action set $A_i$, and the state transitions according to the stochastic function $\mathcal{T}: \mathcal{S} \times \mathcal{A}^1 \times \dots \times \mathcal{A}^N \to \triangle(\mathcal{S})$ represents the set of probability distributions over $\mathcal{S}$. Each agent then receives a reward based on the reward function $\mathcal{R}: \mathcal{S} \times \mathcal{A} \rightarrow \mathbb{R}$. The objective of agent $i$ is to learn a policy $\pi^i:\mathcal{O}^i\to\triangle(A^i)$ based on its own observation $o^i=\mathbb{O}(s,i)$ and current reward $r^i(s,a^1,\dotsc,a^N)$ where $s\in S$ represents the current environment state. The goal is to maximize a long-term $\gamma$-discounted payoff under the joint policy $\vec{\pi} = (\pi^1,\dotsc, \pi^N)$ from an initial state $s_0$:
\begin{equation}
V^i_{\vec{\pi}}(s_0)=\mathbb{E}\left[\sum^{T}_{t=0}\gamma^t r^i(s_t,\vec{a}_t) | {\vec{a_t}\sim \vec{\pi}_t,s_{t+1}\sim\mathcal{T}(s_a,\vec{a}_t)}\right],
\end{equation}
where $\gamma\in[0,1]$ is the discount factor, and $T$ denotes the time horizon. We extend POMGS by incorporating reputation assessments, formalized as the function $p_{\eta^i_t}:\tau^i_t \to \{0,1\}^{|\Omega^i|}$ where $\Omega^i$ represents the neighbour set for agent $i$ and $\tau^i_t\coloneq\bigcup_{j\in\Omega^i}((\mathcal{S\times A}^j)_t\times\mathcal{S})$ captures state-action trajectories of $i$ with its neighbours at timestep $t$. Accordingly, $\tau_t\coloneq\bigcup_{i\in N}\tau^i_t$ aggregates trajectories across all agents. In other words, the reputation assessment $p^i$ evaluates the goodness of each neighbour, assigning $1$ for \enquote{good} behaviour and $0$ for \enquote{bad} behaviour. We refer to this framework as \textit{reputation-augmented Mar- kov games}, which provides a basis for reputation assignment and coordination in sequential decision-making. The learning processes of the reputation evaluation and the reputation-augmented reward function will be detailed in the following section.

\section{Methodology}

\begin{figure*}[h]
\centering
  \includegraphics[width=\textwidth,keepaspectratio]{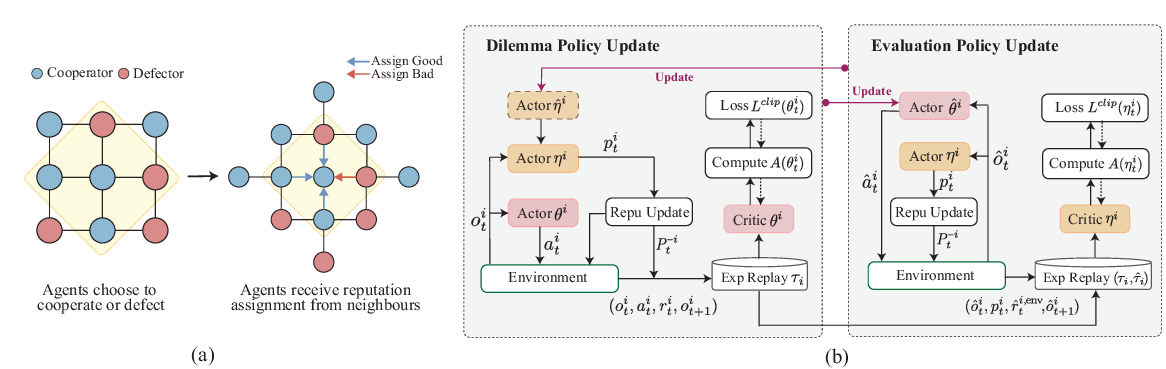}
  \caption{Overview of the social dilemma game with reputation and architecture of our LR2 agents. \normalfont(a) Each agent is connected to four neighbours in a network. Each round consists of two phases: First, agents choose to cooperate or defect based on their reputations and those of their surrounding neighbours.  In the second phase, agents receive reputation assignment reflecting how their behaviours are perceived within their neighbours' local group. (b) Agent $i$ updates its dilemma policy $\theta^i$ by considering both environmental rewards and assigned reputation. The evaluation policy $\eta^i$ is then updated based on the rewards accumulated by the updated dilemma policy $\hat{\eta}^i$.}
  \Description{Overview of the social dilemma game with reputation and architecture of our LR2 agent.}
  \label{fig: overview_framework}
\end{figure*}

Here, we introduce the Learning with Reputation Reward (LR2) met- hod, which allows agents to learn an evaluation function by explicitly accounting for their impact on neighbours and their own objectives. As shown in Figure~\ref{fig: overview_framework}a, each agent decides whether to cooperate or defect with its neighbours and then assigns assessments based on local interactions. A detailed LR2 framework is provided below, with the algorithm listed in Appendix \ref{Appendix:Algorithm}.

\subsection{Learning with Reputation Reward}
\label{Sec: Method_Defination}
The core idea of our LR2 approach is to enable agents to learn an evaluation policy that progressively reshapes the rewards of nearby agents by considering the consequences of interactions on their own objectives. This reshaping encourages agents to account for the effects of their behaviour on their neighbours in a more cooperative manner. In the implementation, agent $i$ learns both a dilemma policy and an evaluation policy, parameterised by $\theta^i \in \mathbb{R}^N$ and $\eta^i \in \mathbb{R}^N$, respectively, in different sequences. Let $-i$ denote the set of all neighbours of agent $i$. As shown in Figure \ref{fig: overview_framework}b, LR2 agents optimize their dilemma policies by considering their actions on neighbours, while adjusting evaluation policies to influence neighbour behaviour. Each agent occupies a specific spatial coordinate and interacts only within its von Neumann neighbourhood~\cite{szabo2007evolutionary}. At each timestep $t$, agents engage in multiple pairwise social dilemma games involving strategy selection and reputation assignment.

In the first phase, agent $i$ selects either cooperation or defection as its dilemma strategy for the next round, according to $a_t^i \sim \pi_{\theta^i}(\cdot|s_t)$. After participating in one round of game pairwise interactions with each neighbour, the agent $i$ receives an environmental reward given by:
\begin{equation}
r^{i,\text{env}}_t=\sum_{j\in \Omega^i}{a^{i}}^{\top}\mathcal{M} a^{j},
\end{equation}
We refer to this aggregate reward as the environmental reward that the agent receives from participating in social dilemma games. In the second phase, the neighbours of agent $i$ assign assessments based on the evaluation function $p_{\eta^i}:\mathcal{O}\times A^{-i}\rightarrow \mathbb{R}^{|\Omega^i|}$, which maps the agent’s observation $o^i$ and the actions of its neighbours to a vector of reputation assessments for all surrounding agents. Finally, the reputation of agent $i$ at the timestep $t$ can be updated according to a running average:
\begin{equation}
    P^i_t= \alpha P^i_{t-1}+\frac{1-\alpha}{|\Omega^i|}\sum_{j\in\Omega^i}p_{\eta^{j}}^i(o^j_t,a^{-j}_t),
\end{equation}
where $\alpha$ is a smoothing parameter that quantifies how quickly the agent's reputation is reshaped by nearby assessments and $p^{i,t}_{\eta_j}$ is the reputation assessment assigned to agent $i$ by its neighbour $j$.

During training, LR2 agents update their dilemma policies by taking into account both environmental rewards and the reputations assigned by their neighbours. They then apply a separate policy gradient to the evaluation policy, based on the rewards generated by the updated dilemma actions (see Figure \ref{fig: overview_framework}b). Following the implementation in~\cite{yang2020learning}, we structure the agents update procedure in an online cross-validation manner~\cite{sutton1999policy}, to account for the fact that reputation assessments toward neighbours have a measurable effect only after those neighbours have updated their dilemma policies. We next describe the LR2 method in a step-by-step process below.

\subsection{Dilemma Policy Update}
\label{Sec: Method_Dilemma}
An LR2 learner optimizes its dilemma policy by considering the additional effect of reputation assessments received from neighbouring agents. Instead of optimizing the expected return under current dilemma parameters $V^i(\theta^i,\theta^{-i})$, LR2 optimizes \ $V^i(\theta^i,\theta^{-i},\eta^{-i}$), wh- ich accounts for the neighbours' evaluation policies $\eta^{-i}$. Specifically, it modifies the immediate reward of an agent so that, at each timestep $t$, the aggregate reward of agent $i$ is divided into two components:
\begin{equation}
    r^i(s_t,a_t,\eta^{-i})\coloneq\beta r^{i,\text{env}}(s_t,a_t)+P^i_{t}(1-\beta) r^{i,\text{env}}(s_t,a_t),
    \label{eq:reputation_reward}
\end{equation}
where $\beta$ is a weighting parameter that balances the influence of the environmental reward and the reputation-based reward. This formulation enables agents to adapt their strategies based not only on direct payoffs but also by accounting for how their actions affect their reputation among neighbours, effectively incorporating the influence of neighbouring agents' assessments into their own policy optimization. When $\beta=1$, agents degenerate into selfish entities,  learning independently and driven solely by their own interests.

Given this formulation, the reshaped reward $r_t^i$ depends on the average reputation derived from the neighbour assessments vector $p_{\eta^{-i}}$. Accordingly, agent $i$ learns a dilemma policy $\pi^i$, parameterised by $\theta^i$, to maximize the objective:
\begin{equation}
    \max_{\theta^i} J^{\text{dilemma}}\coloneq\mathbb{E}_\pi \left[\sum^T_{t=0}\gamma^t r^i(s_t,a_t,\eta^{-i}) + \omega \mathcal{H}^{\pi}(\cdot|s_t) \right].
\end{equation}

We incorporate an additional entropy bonus, controlled by the hyperparameter $\omega$, to encourage exploration and mitigate the issue of early convergence~\cite{haarnoja2017reinforcement}. The LR2 agent $i$ updates its policy parameter $\theta^i$ using a policy gradient approach:
\begin{equation}
\label{eq:dilemma_update}
    \hat{\theta}^i\leftarrow \theta^i+\lambda \sum^T_{t=0}\left[\nabla_{\theta^i}\log \pi^i(a^i_t|o^i_t)G^i_t(\tau^i,\eta^{-i})\right],
\end{equation}
where $\lambda$ is the decay learning rate, and $G^i$ represents the discounted return starting from time $t$, which depends on the trajectory $\tau^i$ and the evaluation policies of the neighbours $\eta^{-i}$.

\subsection{Evaluation Policy Update}
\label{Sec: Method_Repu}
With the updated dilemma policy $\pi_{\hat{\theta}}$, the system then generates a new trajectory $\hat{\tau}\coloneq (\hat{s_0},\hat{a}_t, \hat{r}_0,\dotsc, \hat{s_T})$ based on the updated dilemma policy $\pi_{\hat{\theta}}$. To evaluate the quality of their neighbours, agents primarily compare the average performance of their neighbours within their local group. Hence, the evaluation reward $r^{\text{eval}}$ is designed to assess how effectively agent $i$ interacts with a specific neighbour $j$ in comparison to the average interaction with all its neighbours:
\begin{equation}
    r^{i,\text{eval}}(\hat{s_t},\hat{a}_t) \coloneq\sum_{j\in \Omega^i}\left( \hat{r}^{i,j,\text{env}}\left(\hat{s_t},\hat{a}_t\right)-\frac{\hat{r}^{i,\text{env}}\left(\hat{s_t},\hat{a}_t\right)}{|\Omega^i|}\right).
\end{equation}

This comparison allows agents to influence influencing their neighbours' behaviour so as to maximize their own extrinsic rewards by assigning reputation scores to neighbours. Agent $i$ then uses this new trajectory $\hat{\tau}$ to optimize its evaluation policy, parameterised by $\eta^i$, with the following objective:
\begin{equation}
\label{eq:eval_obj}
    \max_{\eta^i} J^{\text{eval}}\coloneq \mathbb{E}_{\hat{\pi}}\left[\sum^T_{t=0}\gamma^t \left(r^{i,\text{eval}}(\hat{s_t},\hat{a}_t)-\mu \mathcal{D}^i(o_t,a_t)\right)\right],
\end{equation}
where $\mathcal{D}^i(o_t,a_t)$ represents the mean-squared-error, controlled by the sensitive parameter $\mu$:
\begin{equation}
    \mathcal{D}^i(o_t,a_t)\coloneq\sum_{j\in \Omega^i}\sum_{k\in \Omega^j} \left[p^j_{\eta^i}(o^i_t,a^{-i}_t)- p^j_{\eta^{k}}(o^k_t,a^{-k}_t)\right]^2.
\end{equation}

Unlike conventional RL settings where penalty terms like behaviour costs are used, we assume that agents have mixed motivations so that they also care about the alignment between their own assessments and others' assessments (i.e., gossip from neighbours' neighbours~\cite{kawakatsu2024mechanistic}) toward the same agent. However, we argue that this negative term incurred by Equation~(\ref{eq:eval_obj}) should not be included in the total reward, as the dilemma policy and evaluation policy are two separate modules. With the new trajectory $\hat{\tau}$, agent $i$ carries out an update regarding the evaluation policy accordingly:
\begin{equation}
    \hat{\eta}^i\leftarrow\eta^i+\lambda f(\hat{\tau}^i,\tau^i,\hat{\theta},\eta^i),
\end{equation}
with the same learning rate used in Equation~(\ref{eq:dilemma_update}). Note that when taking the gradient with respect to $\eta^i$, we need to account for the fact that $\hat{\pi}^{-i}$ depends on $\eta^i$. Finally, in alignment with the loss function proposed in~\cite{yang2020learning}, the update function for the evaluation policy $\pi_\eta^i$ can be expressed as follows:
\begin{equation}
f(\hat{\tau}^i,\tau^i,\hat{\theta},\eta^i)=\sum_{j\in\Omega^i}\sum^T_{t=0}\nabla_{\eta^i}\log\pi_{\hat{\theta^j}}(\hat{a}^j_t|\hat{o}^j_t){G'}^{i}_t(\hat{\tau}^i,\tau,\eta^i), 
\end{equation}
where
\begin{equation} \label{eq:G}
{G'}^i_t(\hat{\tau},\tau,\eta^i)=\sum^T_{l=t}\gamma^{l-t}\left[r^{i,\text{eval}}(\hat{s_t},\hat{a_t})-\mu\mathcal{D}^i(o_t,a_t)\right].
\end{equation}

Here, the first term on the right side of Equation~(\ref{eq:G}) reflects how changes in the agent's evaluation policy $\eta^i$ affect its own expected comparable extrinsic reward $r^{i,\text{eval}}$ through the impact on neighbouring agent's policies $\hat{\theta^j}$. Note that there is no recursive dependence of $\theta^{-i}$ on $\eta^i$ in the second term, as it is included in the previous evaluation episode.



\section{Experimental Setup}
\subsection{Evaluation Domain}
In our experiment, agents are situated on an $L \times L$ square lattice with periodic boundary conditions. Vertices represent agents, and edges indicate relationships with their four neighbours. At the start of each episode, agents are randomly assigned as either cooperators or defectors, with equal probability, and given a random reputation.

In all experiments, we simulate a population of $N=400$, each participating in all episodes. The agent parameters are distributed across $100$ learner processes, each handling policy gradient updates for four agents. To generate experience, $10$ parallel arenas are created, where agents interact with neighbours over $10,000$ episodes of $20$ timesteps each, totalling $200,000$ steps. After each episode, sampled trajectories are aggregated and sent to learner processes for updates. Cooperation levels, representing the average frequency of cooperative strategies, are measured across all scenarios.

\subsection{Practical Algorithm}
To maximize long-term $\gamma$-discounted payoff, a common strategy is for each agent to optimize its policy independently using policy gradient techniques~\cite{sutton1999policy}, such as REINFORCE~\cite{williams1992simple} or Proximal Policy Optimization (PPO)~\cite{schulman2017proximal}. In this paper, we deploy PPO as the learning algorithm. Similar to the asynchronous advantage actor-critic method~\cite{mniha16Asynchronous}, PPO maintains both value (critic) and policy (actor) estimates using deep neural networks and replaces the return $G^i_t$ by $G^i_t-V^i(s_T)$. However, it improves stability by employing a clipped surrogate objective to limit policy updates. We also implement common practices like Generalize Advantage Estimation (GAE)~\cite{Schulmanetal2016High} with advantage normalization and value clipping. Gradients are computed using the Adam optimizer~\cite{kingma2014adam} with a linear annealing learning rate.

\subsection{Design of Network Architectures}
In our implementation, each agent contains two independent neural networks, trained on batches of environmental experiences to update their weights. These policy networks operate separately, formulating policies independently and without sharing parameters across agents. Each network learns abstract representations from observations to take reward-maximizing actions. The architecture consists of a dual-layer perception with $32$ hidden units and employs the activation function $tanh$ for non-linear transformations. 

For the dilemma policy network, the inputs are the received reputation assignments of the agent and its neighbours. The output of the network at each timestep includes a policy containing a probability distribution over the next binary dilemma action (cooperation or defection), and a value function estimating the discounted future return under the current policy. For the evaluation policy network, agents take each neighbour's current dilemma action and the reputation assessments received from all surroundings as the observation input to produce a value function and an updated assessment vector for each neighbour. While all agents learn independently, they coexist within a shared environment where they influence each other's experience and learning.

\section{Results}
To ensure robustness, we average results over the final ten episodes and replicate each experiment five times. In the baseline (D-D) model, agents rely solely on their neighbours' past actions to decide whether to cooperate, without incorporating reputation information. To isolate the contributions of LR2's individual components, we developed ablated versions in which agents train using separate policies (e.g., Independent PPO (IPPO)~\cite{bettini2024benchmarl}), and reputation is decoupled from the reward structure. Although our primary focus is on LR2's performance in the Prisoner's Dilemma (PD) game, we also evaluate its effectiveness in the Stag Hunt (SG) and Snowdrift (SH) games. Finally, we examine the impact of various hyperparameters and the role of reputation-based intrinsic rewards on LR2's ability to promote cooperation in Appendix~\ref{Appendix:hyperparameter}.

\subsection{Effectiveness of Introducing LR2}
\begin{figure}
\centering
  \includegraphics[width=0.5\textwidth,keepaspectratio]{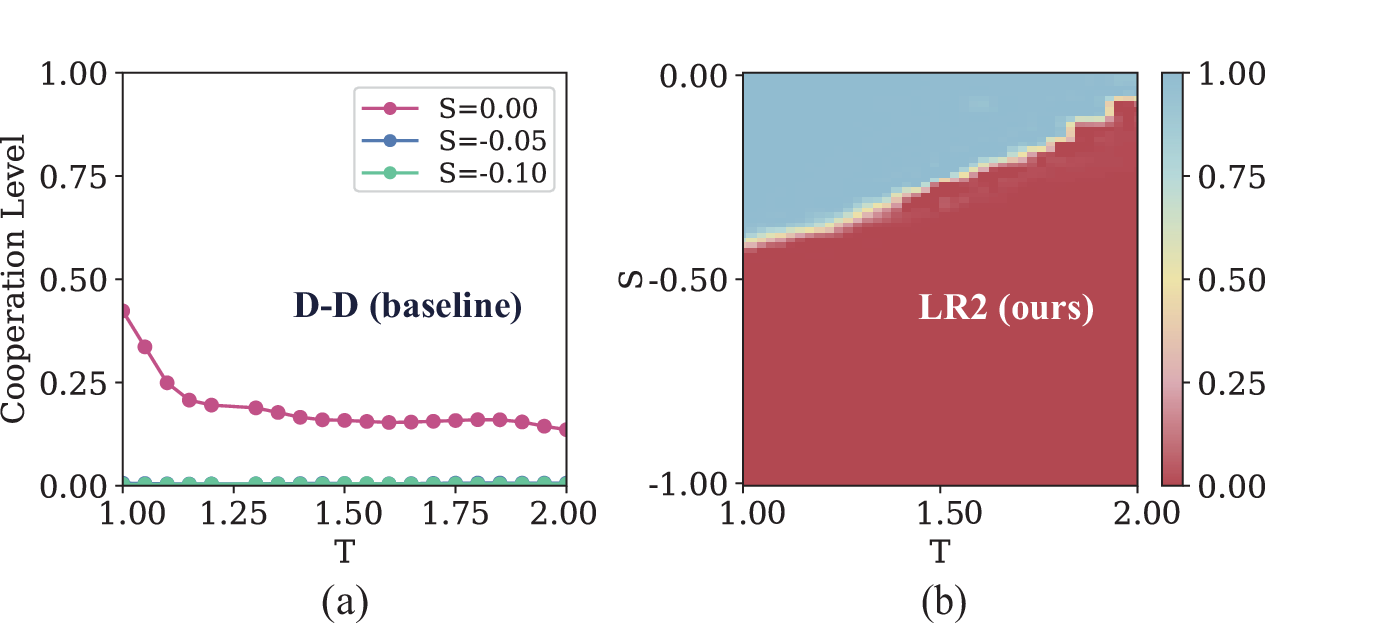}
  \caption{Comparison of cooperation levels between LR2 and the D-D baseline across different $T$ and $S$ values. \normalfont LR2 demonstrates a more effective promotion of cooperation in the PD game. (a) D-D baseline: agents optimize dilemma policies based solely on environmental rewards. (b) LR2 method (ours): agents utilize both dilemma and evaluation policies, with rewards reshaped by reputation. The colour gradient from red to blue represents cooperation levels ranging from $0$ to $1$.}
  \label{fig: heatmap}
\end{figure}
We begin by comparing the performance of our LR2 method against the D-D baseline across various PD scenarios. Figure~\ref{fig: heatmap} depicts cooperation levels as functions of the temptation payoff $T$ and sucker's payoff $S$, with and without reputation assignment and reward reshaping. In Figure~\ref{fig: heatmap}a, we observe that the well-established spatial reciprocity effect in evolutionary game theory~\cite{wang2013insight}—which promotes cooperation through the network structure—fails to materialize in MARL settings using solely dilemma policies: even when exploitation carries no negative consequences (e.g., $S=0$), cooperation remains below $0.25$ in the D-D group. Moreover, as detailed in Appendix~\ref{Appendix:predefined}, predefined social norms do not foster cooperation in structured MARL populations, which is also contrary to recent findings in evolutionary dynamics~\cite{murase2024computational}. In contrast, Figure~\ref{fig: heatmap}b demonstrates that agents using the LR2 method outperform the D-D baseline, significantly expanding the \enquote{wave of cooperation} in the contour plot. Moreover, LR2 agents display high sensitivity to reward changes, with a sharper transition from full cooperation (blue area) to full defection (red area). Interestingly, contrary to previous studies suggesting that cooperation is difficult to sustain under conditions of extreme dilemma strength~\cite{ren2023reputation}, LR2 agents maintain cooperation even when dilemma strength is very high (e.g., $T=2.0$). This suggests the robustness of our approach in promoting the evolution of cooperation. We also assess LR2's robustness under alternative interaction structures in Appendix~\ref{Appendix:network}.

\subsection{Cooperation and Reputation Dynamics}
\begin{figure}
\centering
  \includegraphics[width=0.5\textwidth,keepaspectratio]{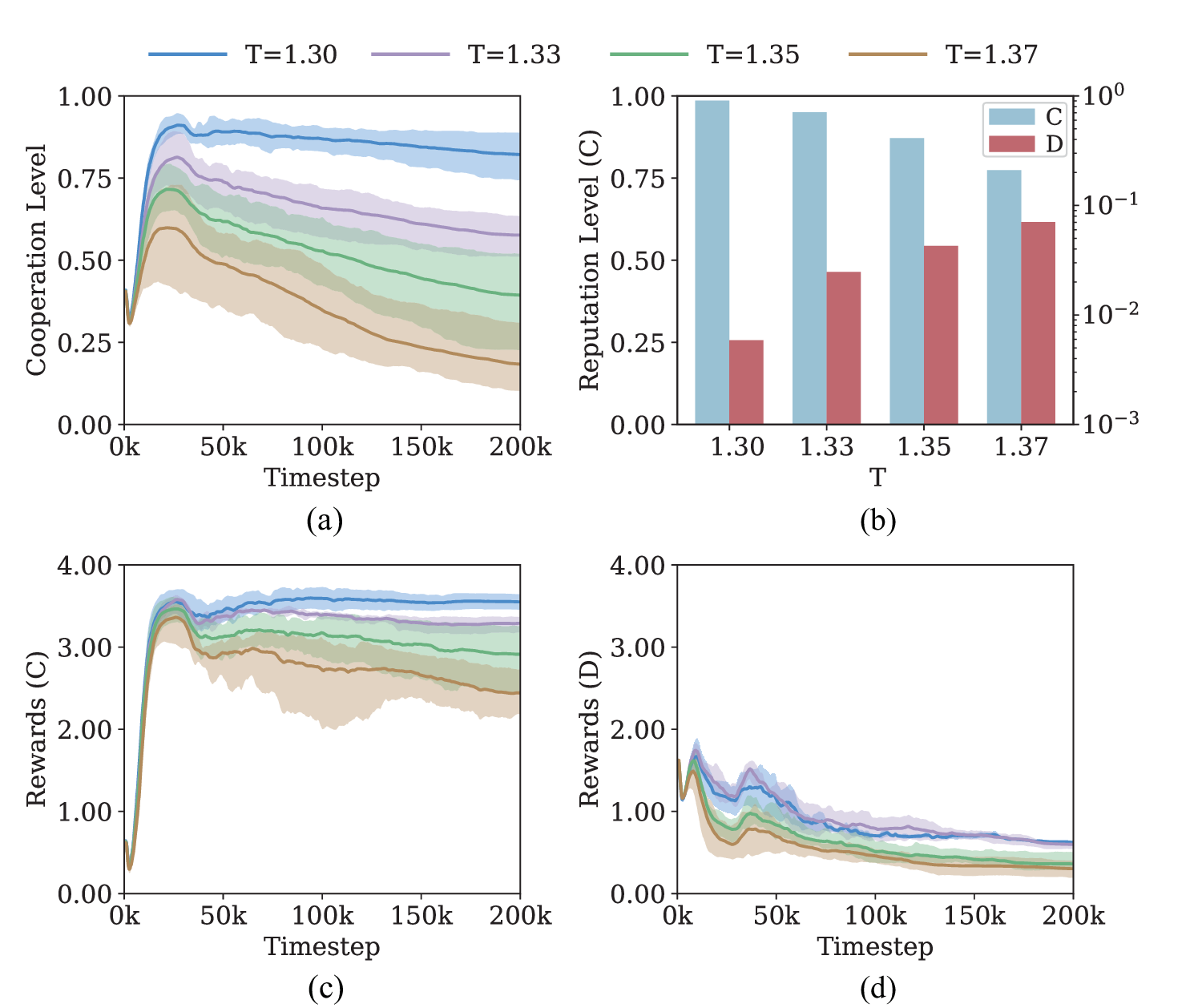}
  \caption{The evolution of cooperation with associated rewards and reputations. \normalfont LR2 agents learn to assess their neighbours' behaviours to reshape rewards, fostering cooperative evolution. The evaluation includes (a) the evolutionary trajectory of cooperation levels, (b) the average reputation of cooperative and defective agents at the end of training, and (c)-(d) the rewards of cooperator and defector over time. Results are presented with the parameter $T$ varying from $1.30$ to $1.37$, while $S$ is fixed at $-0.33$.}
  \label{fig: time_evolution}
\end{figure}
We next illustrate the interplay between the evolution of cooperation and the reputation formulation process, providing a more intuitive explanation of how agents encourage prosocial behaviours by influencing others' learning dynamics. This relationship is detailed in Figure~\ref{fig: time_evolution}, which consists of four subgraphs representing the evolution of cooperation, the corresponding average rewards for cooperative and defective actions, and the reputation distribution for both actions. Each subgraph contains four parameter combinations, reflecting varying dilemma strengths in PD scenarios. As shown in Figure~\ref{fig: time_evolution}a, in all scenarios, the cooperation level exhibits a consistent trend: it briefly declines at the beginning, rises to a peak, and then either stabilizes or gradually decreases. 

\begin{figure*}[t]
\centering
  \includegraphics[width=0.86\textwidth,keepaspectratio]{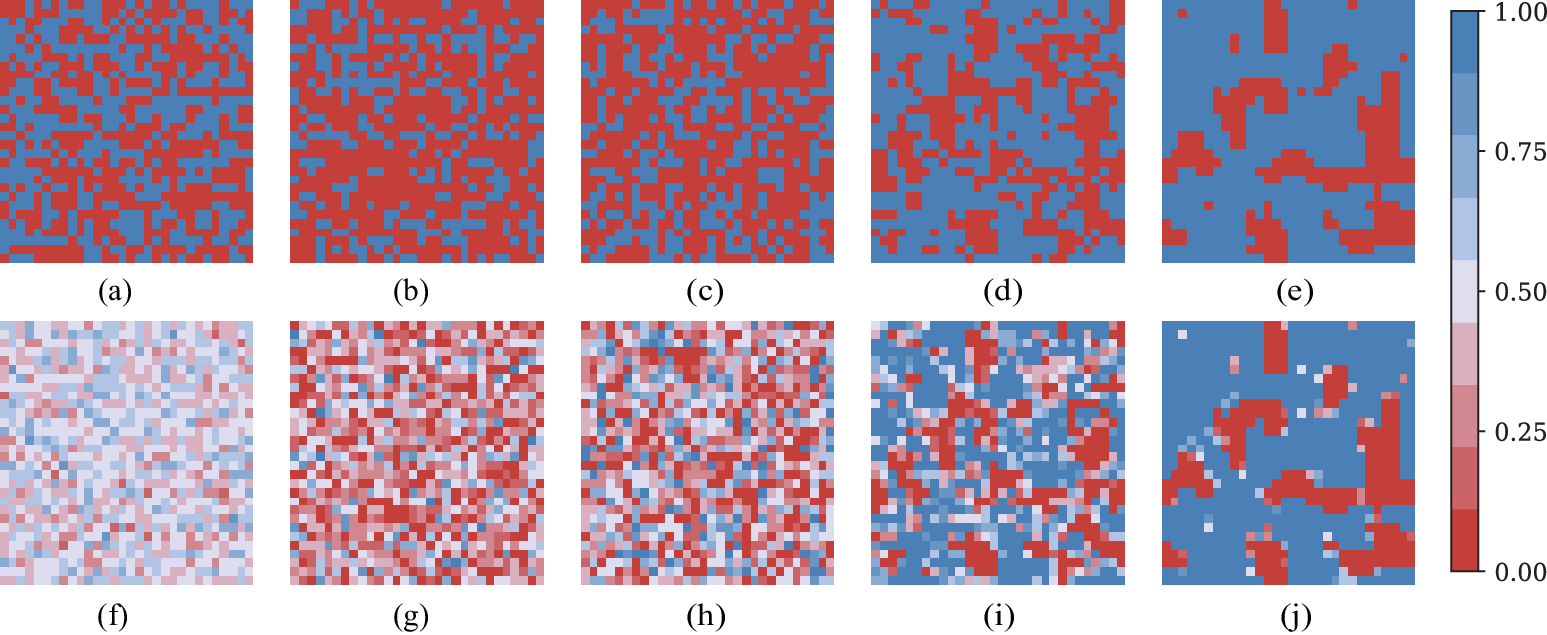}
  \caption{Representative snapshots showing the spatial distribution of learned dilemma actions and assigned reputations on a square lattice. \normalfont The LR2 method fosters the formation of cooperator clusters through spatial effects, thereby enhancing the overall level of cooperation. Panels (a)-(e) display the evolutionary trajectories of two competing dilemma strategies at the timesteps $t= 1$k, $5$k, $10$k, $20$k, and $50$k. The corresponding panels (f)-(j) illustrate the average reputations assigned by neighbours at the same timesteps. Pixels represent agents as cooperators (blue) and defectors (red), with reputation levels ranging from $0$ to $1$. Results are obtained for $T=1.33$ and $S=-0.33$.}
  \Description{Snapshots showing the spatial distribution of learned dilemma actions.}
  \label{fig: strategy_distribution}
\end{figure*}

Since agents' rewards are reshaped by neighbour-assigned reputations, Figure~\ref{fig: time_evolution}b examines the effectiveness of the learned reputation assignment policy by comparing the average reputations of cooperators and defectors at the end of the training process. To highlight the differences, cooperative reputations are shown on a linear scale, while defective ones remain consistently below $0.1$ on a log scale. The group is more sensitive to negative feedback, consistently assigning low reputations to underperforming agents regardless of the dilemma scenario. However, they possess a conservative approach to positive ratings. Notably, even at $T=1.37$, the average reputation of cooperators remains around $0.75$. To further illustrate how these established evaluation strategies guided the LR2 agents' learning, Figures~\ref{fig: time_evolution}c and~\ref{fig: time_evolution}d present the evolutionary trajectories of the average rewards for cooperative and defective actions. Here, the reshaped reward functions as a teaching signal, encouraging agents to act prosocially while discouraging free-riding behaviour. Initially, cooperative rewards are lower than those of defectors, but as agents learn to assign appropriate reputations to their neighbours, the rewards for cooperation gradually surpass those for defection. This demonstrates the synergy between learning reputation evaluations and dilemma-based behaviours. Our findings suggest that effective reputation assignment is crucial; as $T$ increases, agents struggle to identify cooperators, leading to a rapid decline in cooperative rewards and the eventual extinction of cooperators.

\subsection{Formation of Spatial Strategy Patterns}
The above analysis provides an intuitive explanation of the synergy between the established reputation evaluation policy and the evolution of cooperation. To better understand why cooperation emerges within the LR2 framework, Figure~\ref{fig: strategy_distribution} presents characteristic snapshots of agent dilemma action patterns and received reputation assignment, allowing us to further investigate the strategy distribution in the learning dynamics of LR2 agents. Interestingly, we find that LR2 not only reshapes neighbours' payoffs and influences their learning dynamics by effectively identifying dilemma strategies and assigning reputations, but also promotes the formation of cooperative clusters (abbreviated as C-clusters). This spatial pattern formation highlights the evolutionary advantage of the learned evaluation policy, as it helps resist defector invasion and facilitates the emergence of cooperation~\cite{szolnoki2017alliance}. 

Furthermore, the evolutionary path, starting from an initial random state (Figures~\ref{fig: strategy_distribution}a and~\ref{fig: strategy_distribution}f) and progressing to a final equilibrium, can be divided into two distinct phases, consistent with the reported spatial reciprocity effect~\cite{wang2013insight}. The first phase is the enduring period (END), during which cooperators resist the invasion of defectors. This is followed by the second phase called the expansion period (EXP), where the fraction of cooperators begins to increase, signalling the growth and spread of cooperative behaviour. As shown in Figures~\ref{fig: strategy_distribution}(a-b) and~\ref{fig: strategy_distribution}(f-g), during the END period, LR2 agents have not yet learned to effectively evaluate the behaviour of others, leading defectors to avoid reputational damage despite their free-riding behaviour. Consequently, D-clusters expand rapidly, constraining the space available for cooperators. However, once agents learn to assign reputations correctly, they can influence the learning dynamics of others and indirectly protect their own interests. Although cooperation still declines during this phase, the decrease is progressively restrained as C-clusters form incrementally, allowing cooperators to survive and maintain their presence (Figure~\ref{fig: strategy_distribution}c). After a suitable transient period, the population enters the EXP phase, where agents begin to assign reputations accurately. This causes the fragmented D-clusters to gradually disintegrate, while the more cohesive C-clusters expand in size. Moreover, the finding that agents are more sensitive to losses in Figure~\ref{fig: time_evolution}, is confirmed here. As observed in the second row, LR2 agents consistently assign a reputation level of $0$ to defective neighbours but are slower in providing positive evaluations to cooperative behaviours. 

\subsection{Architecture Ablations and Robustness}
\begin{figure}
\centering
  \includegraphics[width=0.50\textwidth,keepaspectratio]{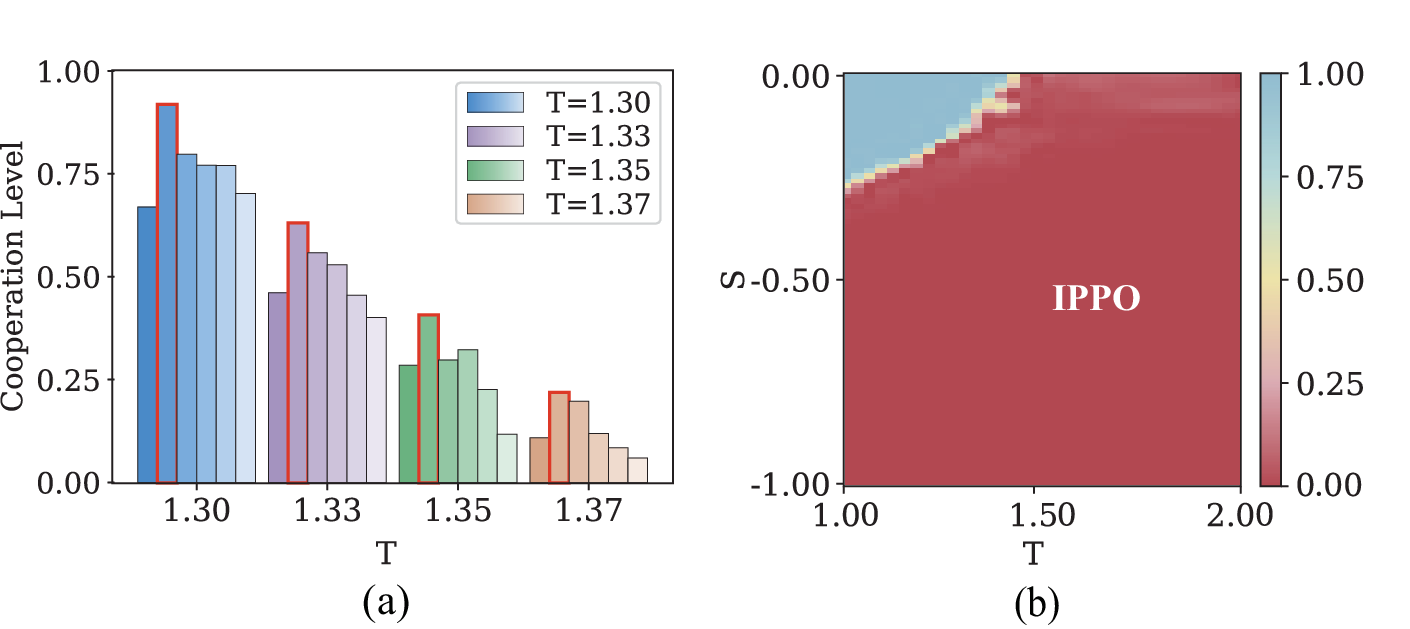}
  \caption{Ablations on LR2 architecture components. \normalfont Considering others' reputation evaluations and the effects of assigned reputations most effectively promotes cooperation.  (a) Reputation alignment varying importance, shown by colours from dark to light representing $\mu$ from $1$ to $0$. Parameter $S$ is fixed at $-0.33$. (b) Performance of the IPPO training method without reputation reward.}
  \label{fig: combine_ablation}
\end{figure}
To better understand LR2, we modified its architecture to address two questions: (1) Does LR2's performance depend on agents forming heterogeneous evaluations of the same behaviour? (2) How is performance affected when agents merely observe reputation without integrating it into reward reshaping? To answer these, we conducted ablation experiments to illustrate the effects of reputation alignment and reward on LR2's performance (Figure~\ref{fig: combine_ablation}). Finally, we verified the robustness of the LR2 method in the SG and SH dilemma settings, with the results summarized in Table~\ref{tab: robustness}.

\textit{Dual impact from local assignment.} In LR2, agents consider both the rewards from interactions and the consistency of their evaluations with others. Since reputation assignment is a biased judgment based on private observation, isolated disagreements may arise. Figure~\ref{fig: combine_ablation}a compares six cases, ranging from completely ignoring reputation alignment ($\mu=0$, darkest bar) to treating it as equally important as rewards ($\mu=1$, lightest bar) in $0.2$ intervals. As shown in all cases, moderate consideration of evaluation consistency ($\mu=0.2$, red border) consistently leads to optimal cooperation. While aligning evaluations with neighbours can help agents learn prosocial behaviour, placing too much emphasis on alignment can be detrimental. For example when $T=1.37$, treating alignment and rewards equally results in the worst performance.

\textit{Decoupling reputation from LR2 rewards.} In Figure~\ref{fig: combine_ablation}b, we consider an IPPO ablation where each agent still trains two policies, dilemma and evaluation, but reputation is used only as observational information and does not affect rewards, by setting $\beta=1$. Compared to the rewards in Figure~\ref{fig: heatmap}, IPPO still outperforms the baseline by using reputation to assess the quality of an individual's past behaviour, since it enriches observational information. However, the \enquote{wave of cooperation} shrinks sharply compared to LR2 training. It suggests that reward shaping proves crucial for sustaining cooperative behaviour, particularly when dilemma strength intensifies, as seen in the right top corner of both contour plots.
\begin{table}[h]
  \caption{Comparison among LR2 agents, D-D baseline, and an ablated IPPO method across three types of dilemmas. \normalfont Performance is measured by the average cooperation level after training. Agents trained using the LR2 method (ours) outperform the other two methods in all scenarios.}
  \centering
  \resizebox{0.49\textwidth}{!}{
    \begin{tabular}{lcccccc}
        \toprule
        \multirow{2}*{Method} & \multicolumn{2}{c}{Prisoner's Dilemma} & \multicolumn{2}{c}{Snowdrift Game}  & \multicolumn{2}{c}{Stag-Hunt Game} \\
        \cmidrule(rl){2-3} \cmidrule(rl){4-5} \cmidrule(rl){6-7} 
        ~ & {($1.1$,$-0.1$)} & {($1.3$,$-0.3$)} &  {($1.1$,$0.1$)} & {($1.3$,$0.3$)} & {($0.9$,$-0.1$)} & {($0.7$,$-0.3$)} \\
        \midrule
        D-D & $0.00$ & $0.00$ & $0.50$ & $0.49$ & $0.10$ & $0.28$  \\
        IPPO & $0.95$ & $0.00$ & $0.99$ & $0.98$ & $0.98$ & $0.99$  \\
        \textbf{LR2 (ours)}& $1.00$ & $0.98$ & $1.00$ & $0.99$ & $1.00$ & $1.00$  \\
        \bottomrule
    \end{tabular}
  }
  \label{tab: robustness}
\end{table}

\textit{Robustness of LR2 in different dilemma types.} We conclude the Results section by evaluating the robustness of the proposed LR2 method in promoting cooperation across different dilemma types, comparing it to the baseline D-D method and ablation IPPO version. As shown in Table~\ref{tab: robustness}, we assess two parameter combinations in each of the SD, SG, and SH scenarios, representing weak and strong dilemma strengths. LR2 consistently exhibits strong performance in both cases, highlighting the effectiveness of our approach. While IPPO approaches similar cooperation levels to LR2  under weak dilemmas, it fails to maintain this effect as the dilemma intensity increases, emphasizing the importance of incorporating reputation rewards and assignments into the learning process. 


\section{Conclusion and future work}
We introduce Learning with Reputation Reward (LR2), a method that leverages reshaped rewards based on bottom-up reputation to promote cooperative behaviour. LR2 incorporates reputation assignment into the dilemma policy learning, while agents concurrently learn an evaluation policy that assigns reputations to incentivise prosocial actions within local groups and maximise their extrinsic objectives. Unlike previous IR methods in MARL that often rely on predefined social norms to guide reactive learning, LR2 allows agents to privately assess each other's reputations based on their own interactions and local observations, further optimizing behaviour using assigned reputation information. This approach provides a pathway to predict the coevolutionary dynamics of reputation and cooperation in structured populations.

Our findings show that LR2 not only stabilizes but also enhances cooperation in spatial social dilemmas. LR2 agents are more sensitive to poor performance, allowing them to effectively identify and penalise defective neighbours. Because reshaped rewards force agents to consider the causal impact of their actions on neighbours, they adopt myopic best responses that lead to improved collective performance. This supports the assertion that even when reputations are privately assessed without enforcement from a top-down institution, emerging reputations can still facilitate cooperation through IR methods~\cite{kawakatsu2024mechanistic,kessinger2023evolution,podder2021local}. Moreover, LR2's promotion of cooperative behaviour is not solely due to reward reshaping but also to its ability to foster the clustering of cooperative strategies, further enhancing spatial reciprocity~\cite{wang2013insight,ren2021evolutionary,he2024temporal}.

Our results raise several unresolved issues regarding the bottom-up reputation in cooperative MARL. While LR2 emphasizes the importance of local observations, agents may receive conflicting information from multiple sources with different weights. Future research could explore how agents process such information in structured populations~\cite{wu2020comprehensive} and consider extending LR2 to more complex evaluation frameworks, such as Melting Pot~\cite{leibo2021scalable}.



\begin{acks}
The authors would like to acknowledge our anonymous reviewers for their thoughtful feedback and thank the assistance given by Research IT and the use of the Computational Shared Facility at The University of Manchester.
\end{acks}


\section*{Supplementary Material}
Source code is available at \url{https://github.com/itstyren/LR2}.



\bibliographystyle{ACM-Reference-Format} 
\bibliography{references}


\begin{thebibliography}{59}


\ifx \showCODEN    \undefined \def \showCODEN     #1{\unskip}     \fi
\ifx \showDOI      \undefined \def \showDOI       #1{#1}\fi
\ifx \showISBNx    \undefined \def \showISBNx     #1{\unskip}     \fi
\ifx \showISBNxiii \undefined \def \showISBNxiii  #1{\unskip}     \fi
\ifx \showISSN     \undefined \def \showISSN      #1{\unskip}     \fi
\ifx \showLCCN     \undefined \def \showLCCN      #1{\unskip}     \fi
\ifx \shownote     \undefined \def \shownote      #1{#1}          \fi
\ifx \showarticletitle \undefined \def \showarticletitle #1{#1}   \fi
\ifx \showURL      \undefined \def \showURL       {\relax}        \fi
\providecommand\bibfield[2]{#2}
\providecommand\bibinfo[2]{#2}
\providecommand\natexlab[1]{#1}
\providecommand\showeprint[2][]{arXiv:#2}

\bibitem[\protect\citeauthoryear{Anastassacos, Garc\'{\i}a, Hailes, and Musolesi}{Anastassacos et~al\mbox{.}}{2021}]%
        {anastassacos2021cooperation}
\bibfield{author}{\bibinfo{person}{Nicolas Anastassacos}, \bibinfo{person}{Julian Garc\'{\i}a}, \bibinfo{person}{Stephen Hailes}, {and} \bibinfo{person}{Mirco Musolesi}.} \bibinfo{year}{2021}\natexlab{}.
\newblock \showarticletitle{Cooperation and Reputation Dynamics with Reinforcement Learning}. In \bibinfo{booktitle}{\emph{Proceedings of the 20th International Conference on Autonomous Agents and MultiAgent Systems}} \emph{(\bibinfo{series}{AAMAS '21})}. \bibinfo{address}{Richland, SC}, \bibinfo{pages}{115–123}.
\newblock


\bibitem[\protect\citeauthoryear{Anastassacos, Hailes, and Musolesi}{Anastassacos et~al\mbox{.}}{2020}]%
        {anastassacos2020partner}
\bibfield{author}{\bibinfo{person}{Nicolas Anastassacos}, \bibinfo{person}{Stephen Hailes}, {and} \bibinfo{person}{Mirco Musolesi}.} \bibinfo{year}{2020}\natexlab{}.
\newblock \showarticletitle{Partner selection for the emergence of cooperation in multi-agent systems using reinforcement learning}. In \bibinfo{booktitle}{\emph{Proceedings of the AAAI Conference on Artificial Intelligence}}, Vol.~\bibinfo{volume}{34}. \bibinfo{pages}{7047--7054}.
\newblock


\bibitem[\protect\citeauthoryear{Bettini, Prorok, and Moens}{Bettini et~al\mbox{.}}{2024}]%
        {bettini2024benchmarl}
\bibfield{author}{\bibinfo{person}{Matteo Bettini}, \bibinfo{person}{Amanda Prorok}, {and} \bibinfo{person}{Vincent Moens}.} \bibinfo{year}{2024}\natexlab{}.
\newblock \showarticletitle{Benchmarl: Benchmarking Multi-Agent Reinforcement Learning}.
\newblock \bibinfo{journal}{\emph{Journal of Machine Learning Research}} \bibinfo{volume}{25}, \bibinfo{number}{217} (\bibinfo{year}{2024}), \bibinfo{pages}{1--10}.
\newblock


\bibitem[\protect\citeauthoryear{Conitzer and Oesterheld}{Conitzer and Oesterheld}{2023}]%
        {conitzer2023foundations}
\bibfield{author}{\bibinfo{person}{Vincent Conitzer} {and} \bibinfo{person}{Caspar Oesterheld}.} \bibinfo{year}{2023}\natexlab{}.
\newblock \showarticletitle{Foundations of Cooperative AI}. In \bibinfo{booktitle}{\emph{Proceedings of the Thirty-Seventh AAAI Conference on Artificial Intelligence and Thirty-Fifth Conference on Innovative Applications of Artificial Intelligence and Thirteenth Symposium on Educational Advances in Artificial Intelligence}} \emph{(\bibinfo{series}{AAAI'23/IAAI'23/EAAI'23}, Vol.~\bibinfo{volume}{37})}. \bibinfo{pages}{15359--15367}.
\newblock


\bibitem[\protect\citeauthoryear{Du, Leibo, Islam, Willis, and Sunehag}{Du et~al\mbox{.}}{2023}]%
        {du2023review}
\bibfield{author}{\bibinfo{person}{Yali Du}, \bibinfo{person}{Joel~Z Leibo}, \bibinfo{person}{Usman Islam}, \bibinfo{person}{Richard Willis}, {and} \bibinfo{person}{Peter Sunehag}.} \bibinfo{year}{2023}\natexlab{}.
\newblock \showarticletitle{A Review of Cooperation in Multi-Agent Learning}.
\newblock \bibinfo{journal}{\emph{arXiv preprint arXiv:2312.05162}} (\bibinfo{year}{2023}).
\newblock


\bibitem[\protect\citeauthoryear{Fatima, Jennings, and Wooldridge}{Fatima et~al\mbox{.}}{2024}]%
        {fatima2024learning}
\bibfield{author}{\bibinfo{person}{Shaheen Fatima}, \bibinfo{person}{Nicholas~R Jennings}, {and} \bibinfo{person}{Michael Wooldridge}.} \bibinfo{year}{2024}\natexlab{}.
\newblock \showarticletitle{Learning to Resolve Social Dilemmas: A Survey}.
\newblock \bibinfo{journal}{\emph{Journal of Artificial Intelligence Research}}  \bibinfo{volume}{79} (\bibinfo{year}{2024}), \bibinfo{pages}{895--969}.
\newblock


\bibitem[\protect\citeauthoryear{Foerster, Chen, Al-Shedivat, Whiteson, Abbeel, and Mordatch}{Foerster et~al\mbox{.}}{2018}]%
        {foerster2018learning}
\bibfield{author}{\bibinfo{person}{Jakob Foerster}, \bibinfo{person}{Richard~Y. Chen}, \bibinfo{person}{Maruan Al-Shedivat}, \bibinfo{person}{Shimon Whiteson}, \bibinfo{person}{Pieter Abbeel}, {and} \bibinfo{person}{Igor Mordatch}.} \bibinfo{year}{2018}\natexlab{}.
\newblock \showarticletitle{Learning with Opponent-Learning Awareness}. In \bibinfo{booktitle}{\emph{Proceedings of the 17th International Conference on Autonomous Agents and MultiAgent Systems}} \emph{(\bibinfo{series}{AAMAS '18})}. \bibinfo{address}{Richland, SC}, \bibinfo{pages}{122–130}.
\newblock


\bibitem[\protect\citeauthoryear{Haarnoja, Tang, Abbeel, and Levine}{Haarnoja et~al\mbox{.}}{2017}]%
        {haarnoja2017reinforcement}
\bibfield{author}{\bibinfo{person}{Tuomas Haarnoja}, \bibinfo{person}{Haoran Tang}, \bibinfo{person}{Pieter Abbeel}, {and} \bibinfo{person}{Sergey Levine}.} \bibinfo{year}{2017}\natexlab{}.
\newblock \showarticletitle{Reinforcement Learning with Deep Energy-based Policies}. In \bibinfo{booktitle}{\emph{Proceedings of the 34th International Conference on Machine Learning - Volume 70}} \emph{(\bibinfo{series}{ICML'17})}. \bibinfo{pages}{1352–1361}.
\newblock


\bibitem[\protect\citeauthoryear{Haynes, Luck, McBurney, Mahmoud, V{\'\i}tek, and Miles}{Haynes et~al\mbox{.}}{2017}]%
        {haynes2017engineering}
\bibfield{author}{\bibinfo{person}{Chris Haynes}, \bibinfo{person}{Michael Luck}, \bibinfo{person}{Peter McBurney}, \bibinfo{person}{Samhar Mahmoud}, \bibinfo{person}{Tom{\'a}{\v{s}} V{\'\i}tek}, {and} \bibinfo{person}{Simon Miles}.} \bibinfo{year}{2017}\natexlab{}.
\newblock \showarticletitle{Engineering the Emergence of Norms: A Review}.
\newblock \bibinfo{journal}{\emph{The Knowledge Engineering Review}}  \bibinfo{volume}{32} (\bibinfo{year}{2017}), \bibinfo{pages}{e18}.
\newblock


\bibitem[\protect\citeauthoryear{He, Ren, Zeng, Liang, Yu, and Zheng}{He et~al\mbox{.}}{2024}]%
        {he2024temporal}
\bibfield{author}{\bibinfo{person}{Yujie He}, \bibinfo{person}{Tianyu Ren}, \bibinfo{person}{Xiao-Jun Zeng}, \bibinfo{person}{Huawen Liang}, \bibinfo{person}{Liukai Yu}, {and} \bibinfo{person}{Junjun Zheng}.} \bibinfo{year}{2024}\natexlab{}.
\newblock \showarticletitle{Temporal interaction and its role in the evolution of cooperation}.
\newblock \bibinfo{journal}{\emph{Physical Review E}} \bibinfo{volume}{110}, \bibinfo{number}{2} (\bibinfo{year}{2024}), \bibinfo{pages}{024210}.
\newblock


\bibitem[\protect\citeauthoryear{Hughes, Leibo, Phillips, Tuyls, Due{\~n}ez-Guzman, Garc{\'\i}a~Casta{\~n}eda, Dunning, Zhu, McKee, Koster, et~al\mbox{.}}{Hughes et~al\mbox{.}}{2018}]%
        {hughes2018inequity}
\bibfield{author}{\bibinfo{person}{Edward Hughes}, \bibinfo{person}{Joel~Z Leibo}, \bibinfo{person}{Matthew Phillips}, \bibinfo{person}{Karl Tuyls}, \bibinfo{person}{Edgar Due{\~n}ez-Guzman}, \bibinfo{person}{Antonio Garc{\'\i}a~Casta{\~n}eda}, \bibinfo{person}{Iain Dunning}, \bibinfo{person}{Tina Zhu}, \bibinfo{person}{Kevin McKee}, \bibinfo{person}{Raphael Koster}, {et~al\mbox{.}}} \bibinfo{year}{2018}\natexlab{}.
\newblock \showarticletitle{Inequity Aversion Improves Cooperation in Intertemporal Social Dilemmas}.
\newblock \bibinfo{journal}{\emph{Advances in Neural Information Processing Systems}}  \bibinfo{volume}{31} (\bibinfo{year}{2018}).
\newblock


\bibitem[\protect\citeauthoryear{Kawakatsu, Kessinger, and Plotkin}{Kawakatsu et~al\mbox{.}}{2024}]%
        {kawakatsu2024mechanistic}
\bibfield{author}{\bibinfo{person}{Mari Kawakatsu}, \bibinfo{person}{Taylor~A Kessinger}, {and} \bibinfo{person}{Joshua~B Plotkin}.} \bibinfo{year}{2024}\natexlab{}.
\newblock \showarticletitle{A Mechanistic Model of Gossip, Reputations, and Cooperation}.
\newblock \bibinfo{journal}{\emph{Proceedings of the National Academy of Sciences}} \bibinfo{volume}{121}, \bibinfo{number}{20} (\bibinfo{year}{2024}), \bibinfo{pages}{e2400689121}.
\newblock


\bibitem[\protect\citeauthoryear{Kessinger, Tarnita, and Plotkin}{Kessinger et~al\mbox{.}}{2023}]%
        {kessinger2023evolution}
\bibfield{author}{\bibinfo{person}{Taylor~A Kessinger}, \bibinfo{person}{Corina~E Tarnita}, {and} \bibinfo{person}{Joshua~B Plotkin}.} \bibinfo{year}{2023}\natexlab{}.
\newblock \showarticletitle{Evolution of Norms for Judging Social Behavior}.
\newblock \bibinfo{journal}{\emph{Proceedings of the National Academy of Sciences}} \bibinfo{volume}{120}, \bibinfo{number}{24} (\bibinfo{year}{2023}), \bibinfo{pages}{e2219480120}.
\newblock


\bibitem[\protect\citeauthoryear{Kingma}{Kingma}{2015}]%
        {kingma2014adam}
\bibfield{author}{\bibinfo{person}{Diederik~P Kingma}.} \bibinfo{year}{2015}\natexlab{}.
\newblock \showarticletitle{Adam: A method for Stochastic Optimization}. In \bibinfo{booktitle}{\emph{Proceedings of the International Conference on Learning Representations (ICLR)}}.
\newblock


\bibitem[\protect\citeauthoryear{Leibo, Due{\~n}ez-Guzman, Vezhnevets, Agapiou, Sunehag, Koster, Matyas, Beattie, Mordatch, and Graepel}{Leibo et~al\mbox{.}}{2021}]%
        {leibo2021scalable}
\bibfield{author}{\bibinfo{person}{Joel~Z Leibo}, \bibinfo{person}{Edgar~A Due{\~n}ez-Guzman}, \bibinfo{person}{Alexander Vezhnevets}, \bibinfo{person}{John~P Agapiou}, \bibinfo{person}{Peter Sunehag}, \bibinfo{person}{Raphael Koster}, \bibinfo{person}{Jayd Matyas}, \bibinfo{person}{Charlie Beattie}, \bibinfo{person}{Igor Mordatch}, {and} \bibinfo{person}{Thore Graepel}.} \bibinfo{year}{2021}\natexlab{}.
\newblock \showarticletitle{Scalable Evaluation of Multi-Agent Reinforcement Learning with Melting Pot}. In \bibinfo{booktitle}{\emph{Proceedings of the 38th International Conference on International Conference on Machine Learning - Volume 139}} \emph{(\bibinfo{series}{ICML'21})}. \bibinfo{pages}{6187--6199}.
\newblock


\bibitem[\protect\citeauthoryear{Leibo, Zambaldi, Lanctot, Marecki, and Graepel}{Leibo et~al\mbox{.}}{2017}]%
        {leibo2017multi}
\bibfield{author}{\bibinfo{person}{Joel~Z. Leibo}, \bibinfo{person}{Vinicius Zambaldi}, \bibinfo{person}{Marc Lanctot}, \bibinfo{person}{Janusz Marecki}, {and} \bibinfo{person}{Thore Graepel}.} \bibinfo{year}{2017}\natexlab{}.
\newblock \showarticletitle{Multi-agent Reinforcement Learning in Sequential Social Dilemmas}. In \bibinfo{booktitle}{\emph{Proceedings of the 16th Conference on Autonomous Agents and MultiAgent Systems}} \emph{(\bibinfo{series}{AAMAS '17})}. \bibinfo{address}{Richland, SC}, \bibinfo{pages}{464–473}.
\newblock


\bibitem[\protect\citeauthoryear{Liang, Liu, Chen, Liu, and Yang}{Liang et~al\mbox{.}}{2022}]%
        {liang2022federated}
\bibfield{author}{\bibinfo{person}{Xinle Liang}, \bibinfo{person}{Yang Liu}, \bibinfo{person}{Tianjian Chen}, \bibinfo{person}{Ming Liu}, {and} \bibinfo{person}{Qiang Yang}.} \bibinfo{year}{2022}\natexlab{}.
\newblock \showarticletitle{Federated Transfer Reinforcement Learning for Autonomous Driving}.
\newblock In \bibinfo{booktitle}{\emph{Federated and Transfer Learning}}. \bibinfo{publisher}{Springer}, \bibinfo{pages}{357--371}.
\newblock


\bibitem[\protect\citeauthoryear{Littman}{Littman}{1994}]%
        {littman1994markov}
\bibfield{author}{\bibinfo{person}{Michael~L. Littman}.} \bibinfo{year}{1994}\natexlab{}.
\newblock \showarticletitle{Markov Games as A Framework for Multi-Agent Reinforcement Learning}.
\newblock In \bibinfo{booktitle}{\emph{Machine Learning Proceedings 1994}}. \bibinfo{publisher}{Elsevier}, \bibinfo{pages}{157--163}.
\newblock


\bibitem[\protect\citeauthoryear{Liu, Szepesv\'{a}ri, and Jin}{Liu et~al\mbox{.}}{2022}]%
        {liu2022sample}
\bibfield{author}{\bibinfo{person}{Qinghua Liu}, \bibinfo{person}{Csaba Szepesv\'{a}ri}, {and} \bibinfo{person}{Chi Jin}.} \bibinfo{year}{2022}\natexlab{}.
\newblock \showarticletitle{Sample-Efficient Reinforcement Learning of Partially Observable Markov Games}. In \bibinfo{booktitle}{\emph{Proceedings of the 36th International Conference on Neural Information Processing Systems}} \emph{(\bibinfo{series}{NIPS '22}, Vol.~\bibinfo{volume}{35})}. \bibinfo{address}{Red Hook, NY, USA}, \bibinfo{pages}{18296--18308}.
\newblock


\bibitem[\protect\citeauthoryear{Lowe, Wu, Tamar, Harb, Abbeel, and Mordatch}{Lowe et~al\mbox{.}}{2017}]%
        {lowe2017multi}
\bibfield{author}{\bibinfo{person}{Ryan Lowe}, \bibinfo{person}{Yi Wu}, \bibinfo{person}{Aviv Tamar}, \bibinfo{person}{Jean Harb}, \bibinfo{person}{Pieter Abbeel}, {and} \bibinfo{person}{Igor Mordatch}.} \bibinfo{year}{2017}\natexlab{}.
\newblock \showarticletitle{Multi-Agent Actor-Critic for Mixed Cooperative-Competitive Environments}. In \bibinfo{booktitle}{\emph{Proceedings of the 31st International Conference on Neural Information Processing Systems}} \emph{(\bibinfo{series}{NIPS'17})}. \bibinfo{address}{Red Hook, NY, USA}, \bibinfo{pages}{6382–6393}.
\newblock


\bibitem[\protect\citeauthoryear{Macy and Flache}{Macy and Flache}{2002}]%
        {macy2002learning}
\bibfield{author}{\bibinfo{person}{Michael~W. Macy} {and} \bibinfo{person}{Andreas Flache}.} \bibinfo{year}{2002}\natexlab{}.
\newblock \showarticletitle{Learning Dynamics in Social Dilemmas}.
\newblock \bibinfo{journal}{\emph{Proceedings of the National Academy of Sciences}} \bibinfo{volume}{99}, \bibinfo{number}{suppl\_3} (\bibinfo{year}{2002}), \bibinfo{pages}{7229--7236}.
\newblock


\bibitem[\protect\citeauthoryear{McKee, Hughes, Zhu, Chadwick, Koster, Castaneda, Beattie, Graepel, Botvinick, and Leibo}{McKee et~al\mbox{.}}{2021}]%
        {mckee2021multi}
\bibfield{author}{\bibinfo{person}{Kevin~R McKee}, \bibinfo{person}{Edward Hughes}, \bibinfo{person}{Tina~O Zhu}, \bibinfo{person}{Martin~J Chadwick}, \bibinfo{person}{Raphael Koster}, \bibinfo{person}{Antonio~Garcia Castaneda}, \bibinfo{person}{Charlie Beattie}, \bibinfo{person}{Thore Graepel}, \bibinfo{person}{Matt Botvinick}, {and} \bibinfo{person}{Joel~Z Leibo}.} \bibinfo{year}{2021}\natexlab{}.
\newblock \showarticletitle{A Multi-Agent Reinforcement Learning Model of Reputation and Cooperation in Human Groups}.
\newblock \bibinfo{journal}{\emph{arXiv preprint arXiv:2103.04982}} (\bibinfo{year}{2021}).
\newblock


\bibitem[\protect\citeauthoryear{McKee, Tacchetti, Bakker, Balaguer, Campbell-Gillingham, Everett, and Botvinick}{McKee et~al\mbox{.}}{2023}]%
        {mckee2023scaffolding}
\bibfield{author}{\bibinfo{person}{Kevin~R McKee}, \bibinfo{person}{Andrea Tacchetti}, \bibinfo{person}{Michiel~A Bakker}, \bibinfo{person}{Jan Balaguer}, \bibinfo{person}{Lucy Campbell-Gillingham}, \bibinfo{person}{Richard Everett}, {and} \bibinfo{person}{Matthew Botvinick}.} \bibinfo{year}{2023}\natexlab{}.
\newblock \showarticletitle{Scaffolding Cooperation in Human Groups with Deep Reinforcement Learning}.
\newblock \bibinfo{journal}{\emph{Nature Human Behaviour}} \bibinfo{volume}{7}, \bibinfo{number}{10} (\bibinfo{year}{2023}), \bibinfo{pages}{1787--1796}.
\newblock


\bibitem[\protect\citeauthoryear{Mnih, Badia, Mirza, Graves, Harley, Lillicrap, Silver, and Kavukcuoglu}{Mnih et~al\mbox{.}}{2016}]%
        {mniha16Asynchronous}
\bibfield{author}{\bibinfo{person}{Volodymyr Mnih}, \bibinfo{person}{Adri\`{a}~Puigdom\`{e}nech Badia}, \bibinfo{person}{Mehdi Mirza}, \bibinfo{person}{Alex Graves}, \bibinfo{person}{Tim Harley}, \bibinfo{person}{Timothy~P. Lillicrap}, \bibinfo{person}{David Silver}, {and} \bibinfo{person}{Koray Kavukcuoglu}.} \bibinfo{year}{2016}\natexlab{}.
\newblock \showarticletitle{Asynchronous Methods for Deep Reinforcement Learning}. In \bibinfo{booktitle}{\emph{Proceedings of the 33rd International Conference on International Conference on Machine Learning - Volume 48}} \emph{(\bibinfo{series}{ICML'16})}. \bibinfo{pages}{1928–1937}.
\newblock


\bibitem[\protect\citeauthoryear{Murase and Hilbe}{Murase and Hilbe}{2024}]%
        {murase2024computational}
\bibfield{author}{\bibinfo{person}{Yohsuke Murase} {and} \bibinfo{person}{Christian Hilbe}.} \bibinfo{year}{2024}\natexlab{}.
\newblock \showarticletitle{Computational evolution of social norms in well-mixed and group-structured populations}.
\newblock \bibinfo{journal}{\emph{Proceedings of the National Academy of Sciences}} \bibinfo{volume}{121}, \bibinfo{number}{33} (\bibinfo{year}{2024}), \bibinfo{pages}{e2406885121}.
\newblock


\bibitem[\protect\citeauthoryear{Nowak}{Nowak}{2006}]%
        {nowak2006five}
\bibfield{author}{\bibinfo{person}{Martin~A Nowak}.} \bibinfo{year}{2006}\natexlab{}.
\newblock \showarticletitle{Five Rules for the Evolution of Cooperation}.
\newblock \bibinfo{journal}{\emph{Science}} \bibinfo{volume}{314}, \bibinfo{number}{5805} (\bibinfo{year}{2006}), \bibinfo{pages}{1560--1563}.
\newblock


\bibitem[\protect\citeauthoryear{Nowak and Sigmund}{Nowak and Sigmund}{2005}]%
        {nowak2005evolution}
\bibfield{author}{\bibinfo{person}{Martin~A Nowak} {and} \bibinfo{person}{Karl Sigmund}.} \bibinfo{year}{2005}\natexlab{}.
\newblock \showarticletitle{Evolution of Indirect Reciprocity}.
\newblock \bibinfo{journal}{\emph{Nature}} \bibinfo{volume}{437}, \bibinfo{number}{7063} (\bibinfo{year}{2005}), \bibinfo{pages}{1291--1298}.
\newblock


\bibitem[\protect\citeauthoryear{Ohtsuki and Iwasa}{Ohtsuki and Iwasa}{2004}]%
        {ohtsuki2004should}
\bibfield{author}{\bibinfo{person}{Hisashi Ohtsuki} {and} \bibinfo{person}{Yoh Iwasa}.} \bibinfo{year}{2004}\natexlab{}.
\newblock \showarticletitle{How Should We Define Goodness? —Reputation Dynamics in Indirect Reciprocity}.
\newblock \bibinfo{journal}{\emph{Journal of Theoretical Biology}} \bibinfo{volume}{231}, \bibinfo{number}{1} (\bibinfo{year}{2004}), \bibinfo{pages}{107--120}.
\newblock


\bibitem[\protect\citeauthoryear{Oldenburg and Zhi-Xuan}{Oldenburg and Zhi-Xuan}{2024}]%
        {oldenburg2024learning}
\bibfield{author}{\bibinfo{person}{Ninell Oldenburg} {and} \bibinfo{person}{Tan Zhi-Xuan}.} \bibinfo{year}{2024}\natexlab{}.
\newblock \showarticletitle{Learning and Sustaining Shared Normative Systems via Bayesian Rule Induction in Markov Games}. In \bibinfo{booktitle}{\emph{Proceedings of the 23rd International Conference on Autonomous Agents and Multiagent Systems}} \emph{(\bibinfo{series}{AAMAS '24})}. \bibinfo{address}{Richland, SC}, \bibinfo{pages}{1510–1520}.
\newblock


\bibitem[\protect\citeauthoryear{Perolat, Leibo, Zambaldi, Beattie, Tuyls, and Graepel}{Perolat et~al\mbox{.}}{2017}]%
        {perolat2017multi}
\bibfield{author}{\bibinfo{person}{Julien Perolat}, \bibinfo{person}{Joel~Z. Leibo}, \bibinfo{person}{Vinicius Zambaldi}, \bibinfo{person}{Charles Beattie}, \bibinfo{person}{Karl Tuyls}, {and} \bibinfo{person}{Thore Graepel}.} \bibinfo{year}{2017}\natexlab{}.
\newblock \showarticletitle{A Multi-Agent Reinforcement Learning Model of Common-Pool Resource Appropriation}. In \bibinfo{booktitle}{\emph{Proceedings of the 31st International Conference on Neural Information Processing Systems}} \emph{(\bibinfo{series}{NIPS'17})}. \bibinfo{address}{Red Hook, NY, USA}, \bibinfo{pages}{3646–3655}.
\newblock


\bibitem[\protect\citeauthoryear{Podder, Righi, and Tak{\'a}cs}{Podder et~al\mbox{.}}{2021}]%
        {podder2021local}
\bibfield{author}{\bibinfo{person}{Shirsendu Podder}, \bibinfo{person}{Simone Righi}, {and} \bibinfo{person}{K{\'a}roly Tak{\'a}cs}.} \bibinfo{year}{2021}\natexlab{}.
\newblock \showarticletitle{Local Reputation, Local Selection, and the Leading Eight Norms}.
\newblock \bibinfo{journal}{\emph{Scientific Reports}} \bibinfo{volume}{11}, \bibinfo{number}{1} (\bibinfo{year}{2021}), \bibinfo{pages}{16560}.
\newblock


\bibitem[\protect\citeauthoryear{Ren and Zeng}{Ren and Zeng}{2023}]%
        {ren2023reputation}
\bibfield{author}{\bibinfo{person}{Tianyu Ren} {and} \bibinfo{person}{Xiao-Jun Zeng}.} \bibinfo{year}{2023}\natexlab{}.
\newblock \showarticletitle{Reputation-Based Interaction Promotes Cooperation with Reinforcement Learning}.
\newblock \bibinfo{journal}{\emph{IEEE Transactions on Evolutionary Computation}} \bibinfo{volume}{28}, \bibinfo{number}{1177--1188} (\bibinfo{year}{2023}), \bibinfo{pages}{042145}.
\newblock


\bibitem[\protect\citeauthoryear{Ren and Zeng}{Ren and Zeng}{2024}]%
        {ren2024enhancing}
\bibfield{author}{\bibinfo{person}{Tianyu Ren} {and} \bibinfo{person}{Xiao-Jun Zeng}.} \bibinfo{year}{2024}\natexlab{}.
\newblock \showarticletitle{Enhancing Cooperation through Selective Interaction and Long-term Experiences in Multi-Agent Reinforcement Learning}. In \bibinfo{booktitle}{\emph{Proceedings of the Thirty-Third International Joint Conference on Artificial Intelligence}} \emph{(\bibinfo{series}{IJCAI '24})}. \bibinfo{pages}{193--201}.
\newblock


\bibitem[\protect\citeauthoryear{Ren and Zheng}{Ren and Zheng}{2021}]%
        {ren2021evolutionary}
\bibfield{author}{\bibinfo{person}{Tianyu Ren} {and} \bibinfo{person}{Junjun Zheng}.} \bibinfo{year}{2021}\natexlab{}.
\newblock \showarticletitle{Evolutionary dynamics in the spatial public goods game with tolerance-based expulsion and cooperation}.
\newblock \bibinfo{journal}{\emph{Chaos, Solitons \& Fractals}}  \bibinfo{volume}{151} (\bibinfo{year}{2021}), \bibinfo{pages}{111241}.
\newblock


\bibitem[\protect\citeauthoryear{Santos, Pacheco, and Lenaerts}{Santos et~al\mbox{.}}{2006}]%
        {santos2006evolutionary}
\bibfield{author}{\bibinfo{person}{Francisco~C Santos}, \bibinfo{person}{Jorge~M Pacheco}, {and} \bibinfo{person}{Tom Lenaerts}.} \bibinfo{year}{2006}\natexlab{}.
\newblock \showarticletitle{Evolutionary Dynamics of Social Dilemmas in Structured Heterogeneous Populations}.
\newblock \bibinfo{journal}{\emph{Proceedings of the National Academy of Sciences}} \bibinfo{volume}{103}, \bibinfo{number}{9} (\bibinfo{year}{2006}), \bibinfo{pages}{3490--3494}.
\newblock


\bibitem[\protect\citeauthoryear{Santos, Pacheco, and Santos}{Santos et~al\mbox{.}}{2021}]%
        {santos2021complexity}
\bibfield{author}{\bibinfo{person}{Fernando~P Santos}, \bibinfo{person}{Jorge~M Pacheco}, {and} \bibinfo{person}{Francisco~C Santos}.} \bibinfo{year}{2021}\natexlab{}.
\newblock \showarticletitle{The Complexity of Human Cooperation under Indirect Reciprocity}.
\newblock \bibinfo{journal}{\emph{Philosophical Transactions of the Royal Society B}} \bibinfo{volume}{376}, \bibinfo{number}{1838} (\bibinfo{year}{2021}), \bibinfo{pages}{20200291}.
\newblock


\bibitem[\protect\citeauthoryear{Santos, Santos, and Pacheco}{Santos et~al\mbox{.}}{2016}]%
        {santos2016social}
\bibfield{author}{\bibinfo{person}{Fernando~P Santos}, \bibinfo{person}{Francisco~C Santos}, {and} \bibinfo{person}{Jorge~M Pacheco}.} \bibinfo{year}{2016}\natexlab{}.
\newblock \showarticletitle{Social Norms of Cooperation in Small-Scale Societies}.
\newblock \bibinfo{journal}{\emph{PLOS Computational Biology}} \bibinfo{volume}{12}, \bibinfo{number}{1} (\bibinfo{year}{2016}), \bibinfo{pages}{1--13}.
\newblock


\bibitem[\protect\citeauthoryear{Santos, Santos, and Pacheco}{Santos et~al\mbox{.}}{2018}]%
        {santos2018social}
\bibfield{author}{\bibinfo{person}{Fernando~P Santos}, \bibinfo{person}{Francisco~C Santos}, {and} \bibinfo{person}{Jorge~M Pacheco}.} \bibinfo{year}{2018}\natexlab{}.
\newblock \showarticletitle{Social Norm Complexity and Past Reputations in the Evolution of Cooperation}.
\newblock \bibinfo{journal}{\emph{Nature}} \bibinfo{volume}{555}, \bibinfo{number}{7695} (\bibinfo{year}{2018}), \bibinfo{pages}{242--245}.
\newblock


\bibitem[\protect\citeauthoryear{Savarimuthu and Cranefield}{Savarimuthu and Cranefield}{2011}]%
        {savarimuthu2011norm}
\bibfield{author}{\bibinfo{person}{Bastin Tony~Roy Savarimuthu} {and} \bibinfo{person}{Stephen Cranefield}.} \bibinfo{year}{2011}\natexlab{}.
\newblock \showarticletitle{Norm Creation, Spreading and Emergence: A Survey of Simulation Models of Norms in Multi-Agent Systems}.
\newblock \bibinfo{journal}{\emph{Multiagent and Grid Systems}} \bibinfo{volume}{7}, \bibinfo{number}{1} (\bibinfo{year}{2011}), \bibinfo{pages}{21--54}.
\newblock


\bibitem[\protect\citeauthoryear{Schulman, Moritz, Levine, Jordan, and Abbeel}{Schulman et~al\mbox{.}}{2016}]%
        {Schulmanetal2016High}
\bibfield{author}{\bibinfo{person}{John Schulman}, \bibinfo{person}{Philipp Moritz}, \bibinfo{person}{Sergey Levine}, \bibinfo{person}{Michael Jordan}, {and} \bibinfo{person}{Pieter Abbeel}.} \bibinfo{year}{2016}\natexlab{}.
\newblock \showarticletitle{High-Dimensional Continuous Control Using Generalized Advantage Estimation}. In \bibinfo{booktitle}{\emph{Proceedings of the International Conference on Learning Representations (ICLR)}}.
\newblock


\bibitem[\protect\citeauthoryear{Schulman, Wolski, Dhariwal, Radford, and Klimov}{Schulman et~al\mbox{.}}{2017}]%
        {schulman2017proximal}
\bibfield{author}{\bibinfo{person}{John Schulman}, \bibinfo{person}{Filip Wolski}, \bibinfo{person}{Prafulla Dhariwal}, \bibinfo{person}{Alec Radford}, {and} \bibinfo{person}{Oleg Klimov}.} \bibinfo{year}{2017}\natexlab{}.
\newblock \showarticletitle{Proximal Policy Optimization Algorithms}.
\newblock \bibinfo{journal}{\emph{arXiv preprint arXiv:1707.06347}} (\bibinfo{year}{2017}).
\newblock


\bibitem[\protect\citeauthoryear{Sigmund, De~Silva, Traulsen, and Hauert}{Sigmund et~al\mbox{.}}{2010}]%
        {sigmund2010social}
\bibfield{author}{\bibinfo{person}{Karl Sigmund}, \bibinfo{person}{Hannelore De~Silva}, \bibinfo{person}{Arne Traulsen}, {and} \bibinfo{person}{Christoph Hauert}.} \bibinfo{year}{2010}\natexlab{}.
\newblock \showarticletitle{Social Learning Promotes Institutions for Governing the Commons}.
\newblock \bibinfo{journal}{\emph{Nature}} \bibinfo{volume}{466}, \bibinfo{number}{7308} (\bibinfo{year}{2010}), \bibinfo{pages}{861--863}.
\newblock


\bibitem[\protect\citeauthoryear{Smit and Santos}{Smit and Santos}{2024}]%
        {smit2024learning}
\bibfield{author}{\bibinfo{person}{Martin Smit} {and} \bibinfo{person}{Fernando~P. Santos}.} \bibinfo{year}{2024}\natexlab{}.
\newblock \showarticletitle{Learning Fair Cooperation in Mixed-Motive Games with Indirect Reciprocity}. In \bibinfo{booktitle}{\emph{Proceedings of the Thirty-Third International Joint Conference on Artificial Intelligence}} \emph{(\bibinfo{series}{IJCAI '24})}. \bibinfo{pages}{220--228}.
\newblock


\bibitem[\protect\citeauthoryear{Son, Kim, Kang, Hostallero, and Yi}{Son et~al\mbox{.}}{2019}]%
        {son2019qtran}
\bibfield{author}{\bibinfo{person}{Kyunghwan Son}, \bibinfo{person}{Daewoo Kim}, \bibinfo{person}{Wan~Ju Kang}, \bibinfo{person}{David~Earl Hostallero}, {and} \bibinfo{person}{Yung Yi}.} \bibinfo{year}{2019}\natexlab{}.
\newblock \showarticletitle{QTRAN: Learning to Factorize with Transformation for Cooperative Multi-Agent Reinforcement Learning}. In \bibinfo{booktitle}{\emph{Proceedings of the 33rd International Conference on International Conference on Machine Learning - Volume 97}}. ICML '19, \bibinfo{pages}{5887--5896}.
\newblock


\bibitem[\protect\citeauthoryear{Sunehag, Lever, Gruslys, Czarnecki, Zambaldi, Jaderberg, Lanctot, Sonnerat, Leibo, Tuyls, and Graepel}{Sunehag et~al\mbox{.}}{2018}]%
        {sunehag2018value}
\bibfield{author}{\bibinfo{person}{Peter Sunehag}, \bibinfo{person}{Guy Lever}, \bibinfo{person}{Audrunas Gruslys}, \bibinfo{person}{Wojciech~Marian Czarnecki}, \bibinfo{person}{Vinicius Zambaldi}, \bibinfo{person}{Max Jaderberg}, \bibinfo{person}{Marc Lanctot}, \bibinfo{person}{Nicolas Sonnerat}, \bibinfo{person}{Joel~Z. Leibo}, \bibinfo{person}{Karl Tuyls}, {and} \bibinfo{person}{Thore Graepel}.} \bibinfo{year}{2018}\natexlab{}.
\newblock \showarticletitle{Value-Decomposition Networks for Cooperative Multi-Agent Learning Based on Team Reward}. In \bibinfo{booktitle}{\emph{Proceedings of the 17th International Conference on Autonomous Agents and MultiAgent Systems}} \emph{(\bibinfo{series}{AAMAS '18})}. \bibinfo{address}{Richland, SC}, \bibinfo{pages}{2085–2087}.
\newblock


\bibitem[\protect\citeauthoryear{Sutton, McAllester, Singh, and Mansour}{Sutton et~al\mbox{.}}{1999}]%
        {sutton1999policy}
\bibfield{author}{\bibinfo{person}{Richard~S. Sutton}, \bibinfo{person}{David McAllester}, \bibinfo{person}{Satinder Singh}, {and} \bibinfo{person}{Yishay Mansour}.} \bibinfo{year}{1999}\natexlab{}.
\newblock \showarticletitle{Policy Gradient Methods for Reinforcement Learning with Function Approximation}. In \bibinfo{booktitle}{\emph{Proceedings of the 12th International Conference on Neural Information Processing Systems}} \emph{(\bibinfo{series}{NIPS'99})}. \bibinfo{address}{Cambridge, MA, USA}, \bibinfo{pages}{1057–1063}.
\newblock


\bibitem[\protect\citeauthoryear{Szab{\'o} and Fath}{Szab{\'o} and Fath}{2007}]%
        {szabo2007evolutionary}
\bibfield{author}{\bibinfo{person}{Gy{\"o}rgy Szab{\'o}} {and} \bibinfo{person}{Gabor Fath}.} \bibinfo{year}{2007}\natexlab{}.
\newblock \showarticletitle{Evolutionary Games on Graphs}.
\newblock \bibinfo{journal}{\emph{Physics Reports}} \bibinfo{volume}{446}, \bibinfo{number}{4-6} (\bibinfo{year}{2007}), \bibinfo{pages}{97--216}.
\newblock


\bibitem[\protect\citeauthoryear{Szolnoki and Chen}{Szolnoki and Chen}{2017}]%
        {szolnoki2017alliance}
\bibfield{author}{\bibinfo{person}{Attila Szolnoki} {and} \bibinfo{person}{Xiaojie Chen}.} \bibinfo{year}{2017}\natexlab{}.
\newblock \showarticletitle{Alliance Formation with Exclusion in the Spatial Public Goods Game}.
\newblock \bibinfo{journal}{\emph{Physical Review E}} \bibinfo{volume}{95}, \bibinfo{number}{5} (\bibinfo{year}{2017}), \bibinfo{pages}{052316}.
\newblock


\bibitem[\protect\citeauthoryear{Tennant, Hailes, and Musolesi}{Tennant et~al\mbox{.}}{2023}]%
        {tennant2023modeling}
\bibfield{author}{\bibinfo{person}{Elizaveta Tennant}, \bibinfo{person}{Stephen Hailes}, {and} \bibinfo{person}{Mirco Musolesi}.} \bibinfo{year}{2023}\natexlab{}.
\newblock \showarticletitle{Modeling Moral Choices in Social Dilemmas with Multi-Agent Reinforcement Learning}. In \bibinfo{booktitle}{\emph{Proceedings of the Thirty-Second International Joint Conference on Artificial Intelligence}} \emph{(\bibinfo{series}{IJCAI '23})}. \bibinfo{pages}{317--325}.
\newblock


\bibitem[\protect\citeauthoryear{Van~Lange, Joireman, Parks, and Van~Dijk}{Van~Lange et~al\mbox{.}}{2013}]%
        {van2013psychology}
\bibfield{author}{\bibinfo{person}{Paul~AM Van~Lange}, \bibinfo{person}{Jeff Joireman}, \bibinfo{person}{Craig~D Parks}, {and} \bibinfo{person}{Eric Van~Dijk}.} \bibinfo{year}{2013}\natexlab{}.
\newblock \showarticletitle{The Psychology of Social Dilemmas: A Review}.
\newblock \bibinfo{journal}{\emph{Organizational Behavior and Human Decision Processes}} \bibinfo{volume}{120}, \bibinfo{number}{2} (\bibinfo{year}{2013}), \bibinfo{pages}{125--141}.
\newblock


\bibitem[\protect\citeauthoryear{Van~Veelen, Garc{\'\i}a, Rand, and Nowak}{Van~Veelen et~al\mbox{.}}{2012}]%
        {van2012direct}
\bibfield{author}{\bibinfo{person}{Matthijs Van~Veelen}, \bibinfo{person}{Juli{\'a}n Garc{\'\i}a}, \bibinfo{person}{David~G Rand}, {and} \bibinfo{person}{Martin~A Nowak}.} \bibinfo{year}{2012}\natexlab{}.
\newblock \showarticletitle{Direct Reciprocity in Structured Populations}.
\newblock \bibinfo{journal}{\emph{Proceedings of the National Academy of Sciences}} \bibinfo{volume}{109}, \bibinfo{number}{25} (\bibinfo{year}{2012}), \bibinfo{pages}{9929--9934}.
\newblock


\bibitem[\protect\citeauthoryear{Vinitsky, K{\"o}ster, Agapiou, Du{\'e}{\~n}ez-Guzm{\'a}n, Vezhnevets, and Leibo}{Vinitsky et~al\mbox{.}}{2023}]%
        {vinitsky2023learning}
\bibfield{author}{\bibinfo{person}{Eugene Vinitsky}, \bibinfo{person}{Raphael K{\"o}ster}, \bibinfo{person}{John~P Agapiou}, \bibinfo{person}{Edgar~A Du{\'e}{\~n}ez-Guzm{\'a}n}, \bibinfo{person}{Alexander~S Vezhnevets}, {and} \bibinfo{person}{Joel~Z Leibo}.} \bibinfo{year}{2023}\natexlab{}.
\newblock \showarticletitle{A Learning Agent that Acquires Social Norms from Public Sanctions in Decentralized Multi-Agent Settings}.
\newblock \bibinfo{journal}{\emph{Collective Intelligence}} \bibinfo{volume}{2}, \bibinfo{number}{2} (\bibinfo{year}{2023}), \bibinfo{pages}{26339137231162025}.
\newblock


\bibitem[\protect\citeauthoryear{Wang, Kokubo, Tanimoto, Fukuda, and Shigaki}{Wang et~al\mbox{.}}{2013}]%
        {wang2013insight}
\bibfield{author}{\bibinfo{person}{Zhen Wang}, \bibinfo{person}{Satoshi Kokubo}, \bibinfo{person}{Jun Tanimoto}, \bibinfo{person}{Eriko Fukuda}, {and} \bibinfo{person}{Keizo Shigaki}.} \bibinfo{year}{2013}\natexlab{}.
\newblock \showarticletitle{Insight into the So-Called Spatial Reciprocity}.
\newblock \bibinfo{journal}{\emph{Physical Review E}} \bibinfo{volume}{88}, \bibinfo{number}{4} (\bibinfo{year}{2013}), \bibinfo{pages}{042145}.
\newblock


\bibitem[\protect\citeauthoryear{Williams}{Williams}{1992}]%
        {williams1992simple}
\bibfield{author}{\bibinfo{person}{Ronald~J Williams}.} \bibinfo{year}{1992}\natexlab{}.
\newblock \showarticletitle{Simple Statistical Gradient-Following Algorithms for Connectionist Reinforcement Learning}.
\newblock \bibinfo{journal}{\emph{Machine learning}}  \bibinfo{volume}{8} (\bibinfo{year}{1992}), \bibinfo{pages}{229--256}.
\newblock


\bibitem[\protect\citeauthoryear{Wu, Pan, Chen, Long, Zhang, and Philip}{Wu et~al\mbox{.}}{2020}]%
        {wu2020comprehensive}
\bibfield{author}{\bibinfo{person}{Zonghan Wu}, \bibinfo{person}{Shirui Pan}, \bibinfo{person}{Fengwen Chen}, \bibinfo{person}{Guodong Long}, \bibinfo{person}{Chengqi Zhang}, {and} \bibinfo{person}{S~Yu Philip}.} \bibinfo{year}{2020}\natexlab{}.
\newblock \showarticletitle{A Comprehensive Survey on Graph Neural Networks}.
\newblock \bibinfo{journal}{\emph{IEEE Transactions on Neural Networks and Learning Systems}} \bibinfo{volume}{32}, \bibinfo{number}{1} (\bibinfo{year}{2020}), \bibinfo{pages}{4--24}.
\newblock


\bibitem[\protect\citeauthoryear{Xu, Garc\'{\i}a, and Handfield}{Xu et~al\mbox{.}}{2019}]%
        {xu2019cooperation}
\bibfield{author}{\bibinfo{person}{Jason Xu}, \bibinfo{person}{Julian Garc\'{\i}a}, {and} \bibinfo{person}{Toby Handfield}.} \bibinfo{year}{2019}\natexlab{}.
\newblock \showarticletitle{Cooperation with Bottom-up Reputation Dynamics}. In \bibinfo{booktitle}{\emph{Proceedings of the 18th International Conference on Autonomous Agents and MultiAgent Systems}} \emph{(\bibinfo{series}{AAMAS '19})}. \bibinfo{address}{Richland, SC}, \bibinfo{pages}{269–276}.
\newblock


\bibitem[\protect\citeauthoryear{Yang, Li, Farajtabar, Sunehag, Hughes, and Zha}{Yang et~al\mbox{.}}{2020}]%
        {yang2020learning}
\bibfield{author}{\bibinfo{person}{Jiachen Yang}, \bibinfo{person}{Ang Li}, \bibinfo{person}{Mehrdad Farajtabar}, \bibinfo{person}{Peter Sunehag}, \bibinfo{person}{Edward Hughes}, {and} \bibinfo{person}{Hongyuan Zha}.} \bibinfo{year}{2020}\natexlab{}.
\newblock \showarticletitle{Learning to Incentivize Other Learning Agents}. In \bibinfo{booktitle}{\emph{Proceedings of the 34th International Conference on Neural Information Processing Systems}} \emph{(\bibinfo{series}{NIPS '20})}. \bibinfo{address}{Red Hook, NY, USA}, Article \bibinfo{articleno}{1275}, \bibinfo{numpages}{12}~pages.
\newblock


\bibitem[\protect\citeauthoryear{Yu, Velu, Vinitsky, Gao, Wang, Bayen, and Wu}{Yu et~al\mbox{.}}{2022}]%
        {yu2022surprising}
\bibfield{author}{\bibinfo{person}{Chao Yu}, \bibinfo{person}{Akash Velu}, \bibinfo{person}{Eugene Vinitsky}, \bibinfo{person}{Jiaxuan Gao}, \bibinfo{person}{Yu Wang}, \bibinfo{person}{Alexandre Bayen}, {and} \bibinfo{person}{Yi Wu}.} \bibinfo{year}{2022}\natexlab{}.
\newblock \showarticletitle{The Surprising Effectiveness of PPO in Cooperative Multi-Agent Games}. In \bibinfo{booktitle}{\emph{Proceedings of the 36th International Conference on Neural Information Processing Systems}} \emph{(\bibinfo{series}{NIPS '22}, Vol.~\bibinfo{volume}{35})}. \bibinfo{address}{Red Hook, NY, USA}, \bibinfo{pages}{24611--24624}.
\newblock


\bibitem[\protect\citeauthoryear{Zheng, Trott, Srinivasa, Parkes, and Socher}{Zheng et~al\mbox{.}}{2022}]%
        {zheng2022ai}
\bibfield{author}{\bibinfo{person}{Stephan Zheng}, \bibinfo{person}{Alexander Trott}, \bibinfo{person}{Sunil Srinivasa}, \bibinfo{person}{David~C Parkes}, {and} \bibinfo{person}{Richard Socher}.} \bibinfo{year}{2022}\natexlab{}.
\newblock \showarticletitle{The AI Economist: Taxation Policy Design via Two-Level Deep Multiagent Reinforcement Learning}.
\newblock \bibinfo{journal}{\emph{Science Advances}} \bibinfo{volume}{8}, \bibinfo{number}{18} (\bibinfo{year}{2022}), \bibinfo{pages}{eabk2607}.
\newblock


\end{thebibliography}


\end{document}


\maketitle

\appendix 
\section{Training Details}  \label{Appendix:implementation}
\subsection{Algorithm} 
 \label{Appendix:Algorithm}
Our proposed Learning with Reputation Reward (LR2) framework consists of two interconnected components: the Dilemma Policy ($\pi_{\theta}$) and the Evaluation Policy ($p_{\eta}$). The dilemma policy governs agents' immediate strategic choices by optimizing for environmental rewards while incorporating the influence of reputation assessments. The evaluation policy determines how agents assign reputations to their neighbours based on observed behaviours, shaping future incentives for cooperation or defection. 

This dual-policy structure allows agents to influence their neighbours indirectly via reputation-driven incentives rather than direct learning updates. Through iterative updates, LR2 establishes a feedback loop where reputation acts as a proxy for long-term rewards, promoting sustainable cooperation beyond short-term gains. The following algorithm~\ref{alg: algorithm} outlines LR2's structured workflow, clarifying the role of each equation in its logical flow.

\subsection{Idea behind LR2}  \label{Supp:Hyperparameters}
In our training framework, the environmental reward and reputation update ensure that LR2 agents' decisions reflect both immediate interactions and historical behaviour. The reward aggregates payoffs from neighbours, while the running average reputation update balances both past and recent assessments, maintaining stability and adaptability.

The dilemma policy optimizes decision-making by integrating reputation-reshaped rewards, aligning incentives with group expectations to foster cooperation. Entropy regularization prevents premature convergence, while the policy gradient update incorporates both environmental and reputation-based rewards, reinforcing long-term cooperative behaviour.

The evaluation policy enables agents to assess their neighbours' contributions by comparing individual interactions to the local group average. The evaluation reward ensures that assessments are context-aware, promoting fair reputation assignment. The penalty term encourages consistency between an agent's evaluations and those of its neighbours, enhancing alignment within the group. Lastly, the evaluation policy update captures the impact of reputation adjustments on neighbouring agents' behaviours, ensuring that reputation assignments evolve to reinforce cooperative dynamics effectively.

\begin{algorithm}[H]
    \caption{Learning with Reputation Reward (LR2)}\label{alg: algorithm}
    \begin{algorithmic}
    \FOR{each episode $e = 1$ to $M$} 
        \STATE Initialize policy parameters $\theta^i, \eta^i$ for all agents
        \STATE Set initial reputations $P^i_0$ for all agents
        \FOR{timestep $t=1$ to max-episode-length}
            \FOR{each agent $i$ to $N$}
            \STATE \textbf{Phase 1: Environmental Reward Computation} 
            \STATE Generate trajectories $\tau^i$ using dilemma policy $\pi_{\theta^i}$
            \STATE Compute the environmental reward $r^{i,\text{env}}_t$
            \STATE \textbf{Phase 2: Reputation Assignment} 
            \STATE  Assign reputation $p^{i,t}_{\eta^j}$ to neighbours via $\eta^i$
            \ENDFOR
            \STATE Update agents' reputation using a running average
            \STATE \textbf{Dilemma Policy Update:}
            \FOR{each agent $i$ to $N$}
            \STATE Incorporate the reputation into the reward function
            \STATE Update dilemma policy parameters $\theta^i$ via:
            \begin{equation*}
                \hat{\theta}^i\leftarrow \theta^i+\lambda \sum^T_{t=0}\left[\nabla_{\theta^i}\log \pi^i(a^i_t|o^i_t)G^i_t(\tau^i,\eta^{-i})\right],
            \end{equation*}
            \ENDFOR  
            \STATE \textbf{Evaluation  Policy Update:}
            \FOR{each agent $i$ to $N$}
            \STATE Generate trajectory $\hat{\tau}^i$ using the updated policy $\hat{\theta}^i$
            \STATE Compute evaluation reward $r^{i,\text{eval}}$
            \STATE Update evaluation policy parameters $\eta^i$ via:
            \begin{equation*}
                 \hat{\eta}^i\leftarrow\eta^i+\lambda f(\hat{\tau}^i,\tau^i,\hat{\theta},\eta^i),
            \end{equation*}
            
            \ENDFOR
            
        \ENDFOR
    \ENDFOR
    \end{algorithmic}
\end{algorithm}

\section{Additional Results}  \label{Appendix:additional}
\subsection{Impact of Interaction Structure} \label{Appendix:network}
 
Spatial reciprocity is traditionally recognized as a key mechanism for promoting cooperation by  enabling clusters of cooperative agents to emerge. However, as demonstrated in the main text, MARL agents that optimize individual rewards do not inherently form or sustain such clusters. To further evaluate the robustness of our proposed LR2 method, we extend the analysis to two additional interaction settings: alternative network structures with varying neighbourhood sizes, and well-mixed populations with different group sizes. These experiments depart from the lattice topology used in the main text and aim to demonstrate that LR2's effectiveness is not contingent on a specific interaction structure. 

Table~\ref{tab:robustness} reports the average cooperation levels achieved under six configurations: three spatial settings—the conventional square lattice ($4$ neighbours), the Moore neighbourhood ($8$ neighbours), and the Honeycomb lattice ($3$ neighbours)—and three well-mixed settings with pairwise interactions ($k=1$), small-group interactions ($k=4$), and larger groups ($k=8$). In the square lattice, cooperation decreases from $0.82$ at $T=1.30$ to $0.15$ at $T=1.37$. In the Moore neighbourhood, cooperation is nearly abolished across all $T$ values, whereas the Honeycomb lattice supports higher cooperation ($0.97$ at $T=1.30$ and $0.65$ at $T=1.37$). In well-mixed populations, LR2 maintains near-optimal cooperation in pairwise ($k=1$) and small-group ($k=4$) settings (approximately $0.98–0.99$ across all $T$ values), but cooperation collapses when agents interact with $8$ opponents per round ($k=8$).

\begin{table}[h]
  \caption{Average cooperation levels for LR2 under different interaction configurations.}
  \centering
  \resizebox{0.49\textwidth}{!}{
    \begin{tabular}{lcccc}
        \toprule
       Configuration & \textbf{$T=1.30$} & \textbf{$T=1.33$} & \textbf{$T=1.35$} & \textbf{$T=1.37$}\\
        \midrule
        Lattice& $0.82$ & $0.55$ & $0.33$ & $0.15$  \\
        Moore & $0.01$ & $0.00$ & $0.00$ & $0.00$  \\
        Honeycomb & $0.97$ & $0.80$ & $0.74$ & $0.65$   \\
        Well-mixed(1) & $0.99$ & $0.99$ & $0.99$ & $0.99$ \\
        Well-mixed(4) & $0.99$ & $0.98$ & $0.98$ & $0.98$ \\
        Well-mixed(8) & $0.00$ & $0.00$ & $0.00$ & $0.00$ \\
        \bottomrule
    \end{tabular}
  }
  \label{tab:robustness}
\end{table}

These results underscore the critical role of interaction structure in the evolution of cooperation within MARL settings. LR2 effectively promotes cooperation in environments that facilitate clustering and in well-mixed settings with limited interactions. However, its performance deteriorates in settings with extensive neighbourhood sizes or large well-mixed groups. This analysis reinforces the conclusion that, in the absence of mechanisms to sustain localized interactions, reputation-based reward reshaping may be insufficient to overcome the challenges posed by diluted or overly extensive opponent sets.

\subsection{Additional Hyperparameter Analysis} \label{Appendix:hyperparameter}
We next present supplementary results examining several hyperparameters of the LR2 method. Specifically, we analyse: (i) the sensitivity of the $\beta$ parameter, which modulates agent selfishness; (ii) the impact of entropy weight $\omega$ scheduling on exploration and convergence; and (iii) the robustness of LR2 in the presence of adversarial agents. 

Table~\ref{tab:beta} summarizes the average cooperation levels for different temptation values ($T$) under varying $\beta$ settings. Our experiments indicate that smaller $\beta$ values, which correspond to less selfish behaviour, promote higher levels of cooperation and accelerate learning. Notably, $\beta$ values of $0.5$ yield substantially higher cooperation compared to higher values, particularly under lower temptation scenarios.

\begin{table}[ht] 
    \caption{Sensitivity analysis of $\beta$: Average cooperation levels under different temptation values.} 
    \centering
    \begin{tabular}{lccc} 
    \toprule 
    $T$ & $\beta=0.5$ & $\beta=0.6$ & $\beta=0.7$ \\ 
    \midrule 
    $1.30$ & $0.98$ & $0.82$ & $0.15$ \\ 
    $1.33$ & $0.98$ & $0.55$ & $0.04$ \\ 
    $1.35$ & $0.98$ & $0.33$ & $0.00$ \\ 
    $1.37$ & $0.92$ & $0.15$ & $0.00$ \\ 
    \bottomrule 
    \end{tabular} 
\label{tab:beta} 
\end{table}

In the main text, an annealing schedule is employed to gradually reduce the entropy weight, thereby balancing exploration and convergence. To assess this approach, we compared the annealing schedule against fixed entropy weights ($\omega=0.1$ and $\omega=0.0$). As reported in Table~\ref{tab:entropy}, while a fixed entropy weight of $\omega=0.1$ facilitates exploration, it is insufficient to guide agents toward cooperation under stronger dilemma conditions. In contrast, the annealing strategy supports both adequate exploration in early training and convergence to effective cooperative policies.

\begin{table}[ht] 
    \caption{Impact of entropy weight scheduling on cooperation.} 
    \centering 
    \begin{tabular}{lccc} 
    \toprule $T$ & Annealing & Fixed ($\omega=0.1$) & Fixed ($\omega=0.0$) \\
    \midrule $1.30$ & $0.82$ & $0.80$ & $0.16$ \\
    $1.33$ & $0.55$ & $0.37$ & $0.00$ \\
    $1.35$ & $0.33$ & $0.20$ & $0.00$ \\
    $1.37$ & $0.15$ & $0.16$ & $0.00$ \\ 
    \bottomrule
    \end{tabular} 
\label{tab:entropy}
\end{table}

To evaluate LR2's robustness, we introduced adversarial agents that prioritize environmental rewards over reputation-based intrinsic rewards. In our framework, reputation functions as an intrinsic reward that is shaped and assigned based on neighbouring agents' observations. LR2 guides agents not only to learn how to assign reputation effectively, thereby influencing the behaviour of others but also to increase their own rewards by correctly assigning reputation to steer neighbours' future actions beneficially. When adversarial agents disregard this reputation mechanism in favour of solely environmental rewards, the cooperative signal provided by reputation assignment is disrupted.

\begin{table}[ht] 
    \caption{Average cooperation levels with varying proportions of adversarial agents.} 
    \centering 
    \begin{tabular}{lccc} 
    \toprule $T$ & $100\%$ (LR2) & $90\%$ (LR2) & $70\%$ (LR2) \\ 
    \midrule $1.30$ & $0.82$ & $0.35$ & $0.00$ \\ 
    $1.33$ & $0.55$ & $0.21$ & $0.00$ \\ 
    $1.35$ & $0.33$ & $0.00$ & $0.00$ \\ 
    $1.37$ & $0.15$ & $0.00$ & $0.00$ \\ 
    \bottomrule 
    \end{tabular} 
\label{tab:adversarial} 
\end{table}

Table~\ref{tab:adversarial} presents the cooperation levels when varying the proportion of cooperative LR2 agents. A $100\%$ ratio corresponds to all agents following the LR2 strategy, whereas a $90\%$ ratio indicates that $10\%$ of agents behave adversarially. The results show that even a modest intrusion (i.e., $90\%$ cooperative agents) significantly diminishes overall cooperation, and a further reduction to $70\%$ cooperative agents leads to a complete collapse of cooperative behaviour across all temptation values. It reveals that the reputation mechanism, which not only serves as an intrinsic reward but also allows agents to influence neighbours' future behaviours through accurate reputation assignment, is vulnerable to adversarial behaviour.


\subsection{Benchmarks on Predefined Norms} \label{Appendix:predefined}
To further assess the efficacy of our LR2 method, we incorporate four benchmarks based on predefined social norms for evaluating neighbour behaviour and assigning reputation in a MARL setting. These norms include: Stern Judging (SJ), which assigns a good reputation to a donor who helps a good recipient or refuses help to a bad one, and a bad reputation otherwise. Simple-standing (SS), similar to SJ but more benevolent, SS grants a good reputation to any donor who cooperates, regardless of the recipient's status. Shunning (SH), which is less forgiving than SJ, SH assigns a bad reputation to any donor who defects. Image Score (IS) is a first-order norm where the donor’s action alone determines reputation; cooperation yields a good reputation, while defection results in a bad reputation.

\begin{table}[ht] 
    \caption{Average cooperation levels across different predefined social norms under a MARL setting.} 
    \centering 
    \begin{tabular}{lcccc} 
    \toprule 
    Norm & \textbf{$T=1.30$} & \textbf{$T=1.33$} & \textbf{$T=1.35$} & \textbf{$T=1.37$} \\
    \midrule 
    SJ & $0.00$ & $0.00$ & $0.00$ & $0.00$ \\
    SS & $0.52$ & $0.47$ & $0.42$ & $0.37$ \\
    SH & $0.03$ & $0.00$ & $0.00$ & $0.00$ \\
    IS & $0.80$ & $0.53$ & $0.28$ & $0.11$ \\
    \bottomrule 
    \end{tabular} 
\label{tab:predefined} 
\end{table}

As shown in Table~\ref{tab:predefined}, cooperation levels under these predefined norms are generally lower than those achieved with the LR2 method, particularly under stronger dilemma conditions. Notably, SJ and SH yield near-zero cooperation, while SS and IS offer only moderate improvements. In contrast, LR2 consistently outperforms these benchmarks by learning to assign reputations in a decentralized manner, thereby maximizing individual rewards through effective influence on neighbours' future behaviours and enhancing overall system robustness and adaptability. Moreover, unlike rule-based norms—which require manual design and extensive domain knowledge—our proposed LR2 automatically adapts its reputation assignment to the environment, improving generalization across diverse tasks.